\documentclass[12pt]{article}
\usepackage{epsf,latexsym,amsmath,amssymb,amscd}

\newtheorem{theorem}{Theorem}[section]
\newtheorem{lemma}[theorem]{Lemma}

\newtheorem{corollary}[theorem]{Corollary}

\newtheorem{definition}[theorem]{Definition}

\newcommand{\vs}[1]{\myfield\left( {#1} \right)}
\newcommand{\LR}{{{\rm L}\to{\rm R}}}
\newcommand{\RL}{{{\rm R}\to{\rm L}}}

\newcommand{\trace}{{\rm Tr}}

\newcommand{\myfield}{{\mathbb Q}} 

\newcommand{\objects}[1]{{{\rm Ob}\left( {#1} \right)}} 
\newcommand{\morphisms}[1]{{{\rm Fl}\left( {#1} \right)}} 
\newcommand{\fleches}[2]{{{\rm Fl}^{#1}\left( {#2} \right)}} 
\newcommand{\underfleches}[2]{{{\underline {\rm Fl}}^{#1}\left( {#2} \right)}} 

\newcommand{\twoleftarrows}{\;
  \mbox{\vbox{\hbox{$\leftarrow$}\vskip-.35truecm\hbox{$\leftarrow$}
  \vskip-.05truecm}}\;}
\newcommand{\threeleftarrows}{\;
  \mbox{\vbox{\hbox{$\leftarrow$}\vskip-.35truecm\hbox{$\leftarrow$}
  \vskip-.35truecm\hbox{$\leftarrow$}\vskip-.15truecm}}\;}

\newcommand{\tworightarrows}{\;
  \mbox{\vbox{\hbox{$\rightarrow$}\vskip-.35truecm\hbox{$\rightarrow$}
  \vskip-.05truecm}}\;}
\newcommand{\threerightarrows}{\;
  \mbox{\vbox{\hbox{$\rightarrow$}\vskip-.35truecm\hbox{$\rightarrow$}
  \vskip-.35truecm\hbox{$\rightarrow$}\vskip-.15truecm}}\;}

\newcommand{\kbig}{{K_{\rm big}}}
\newcommand{\ksmall}{{K_{\rm small}}}
\newcommand{\zbig}{{Z_{\rm big}}}
\newcommand{\zsmall}{{Z_{\rm small}}}

\newcommand{\isom}{\simeq} 

\newcommand{\from}{\colon}
\newcommand{\ignore}[1]{}

\newcommand{\espace}{{\em espace \'etal\'e}}
\newcommand{\espaces}{{\em espaces \'etal\'es}}

\newcommand{\Hom}{{\rm Hom}}
\newcommand{\simexp}[2]{{\rm SHom}\left({#1},{#2}\right)}
\newcommand{\rder}{{\underline{\underline{ R}}}}
\newcommand{\lder}{{\underline{\underline{ L}}}}
\newcommand{\cat}[1]{{\Delta_{#1}}}
\newcommand{\dercat}[1]{{\cdb(\myfield({#1}))}}

\newcommand{\ca}{{\cal A}}
\newcommand{\cb}{{\cal B}}
\newcommand{\cd}{{\cal D}}
\newcommand{\cdb}{{\cal D}^{\rm b}}
\newcommand{\cc}{{\cal C}}
\newcommand{\ck}{{\cal K}}

\newcommand{\ce}{{\cal E}}

\newcommand{\ch}{{\cal H}}
\newcommand{\cm}{{\cal M}}
\newcommand{\ci}{{\cal I}}

\newcommand{\cf}{{\cal F}}
\newcommand{\cv}{{\cal V}}
\newcommand{\cp}{{\cal P}}
\newcommand{\cu}{{\cal U}}
\newcommand{\cx}{{\cal X}}
\newcommand{\cy}{{\cal Y}}

\newcommand{\cs}{{\cal S}}

\newcommand{\id}{{\rm Id}}

\newcommand{\reals}{{\mathbb R}}
\newcommand{\integers}{{\mathbb Z}}

\newcommand{\proof}{{\par\noindent {\bf Proof}\space\space}}
\newcommand{\proofbox}{\begin{flushright}$\Box$\end{flushright}}

\title{Cohomology in Grothendieck Topologies and Lower Bounds
in Boolean Complexity}

\author{Joel Friedman\thanks{
        Departments of Computer Science,
        University of British Columbia, Vancouver, BC\ \ V6T 1Z4, CANADA,
	and
        Departments of Mathematics,
        University of British Columbia, Vancouver, BC\ \ V6T 1Z2, CANADA.
        {\tt jf@cs.ubc.ca}, {\tt http://www.math.ubc.ca/\~{}jf}.
        Research supported in part by an NSERC grant.}}

\begin{document}           
\maketitle                 
\begin{abstract}
This paper is motivated by questions such as P vs.\ NP and
other questions in Boolean complexity theory.
We describe an approach to attacking such questions
with cohomology, and we show that using Grothendieck topologies
and other ideas from the Grothendieck school gives new hope for
such an attack.  

We focus on circuit depth complexity, and
consider only finite topological spaces or Grothendieck
topologies based on finite categories; as such, we do not
use algebraic geometry or manifolds.

Given two sheaves on a Grothendieck topology, their
{\em cohomological complexity} is the sum of the dimensions of
their Ext groups.
We seek to model the depth complexity of Boolean
functions by the cohomological complexity of
sheaves on a Grothendieck topology.  
We propose that the logical AND of two Boolean functions
will have its corresponding cohomological complexity
bounded in terms of those of the two functions
using ``virtual zero extensions.''
We propose that the logical
negation of a function
will have its corresponding cohomological complexity
equal to that of the original function using duality theory.
We explain these approaches and show that they are stable under
pullbacks and base change.
It is the subject of
ongoing work to achieve AND and negation bounds simultaneously in a
way that yields an interesting depth lower bound.
\end{abstract}

\section{Introduction}

Over twenty years ago lower bounds for
algebraic decision trees were obtained by counting connected components
(and in principle the sum of the Betti numbers) of associated
topological spaces (see \cite{dobkin,steele,benor}).  
This lead to a hope that problems such as
P~vs.~NP, viewed as lower bound problems in
Boolean circuit complexity of a Boolean function,
could be studied via cohomology, e.g., the sum of the Betti
numbers of a topological space
(associated in some way to the function).  We are unaware of any
essential progress in this direction to date.  (But see \cite{smale}
for a success of algebraic topology and the braid group in
another notion of complexity.)
In fact, there are what might be called ``standard obstacles'' to
this topological approach in Boolean complexity.  In this paper
we show that two obstacles in depth complexity
can be circumvented in a natural way
provided that we (1) generalize the notion of Betti number using
sheaf theory and the derived category, and (2) replace
topological spaces with Grothendieck topologies\footnote{A
Grothendieck topology (by which we mean a ``{\it site},'' 
as in [SGA4.II.1.1.5])
is a generalization of a topological space;
a Grothendieck topology has
just enough structure to define sheaf theory and therefore, cohomology,
and has the properties that, 
roughly speaking, (1) an ``open set'' can be ``included'' in
another (or itself) in more than one way, and (2) the notion of ``local''
or ``refinement'' is not necessarily the ``canonical'' one.},
see [SGA4]\footnote{
Throughout this paper, we use notation such as [SGA4.V.2.3.6] to 
refer to SGA4 (i.e., \cite{sga4.1,sga4.2,sga4.3}), {\em expos\'e} V,
{\em section} (or, in this case, {\em exercice}) 2.3.6.
}.  We explain
our approach to this circumvention, and give some foundational
theorems that we hope will be useful in our ongoing work
of seeking (Grothendieck)
topological models for Boolean functions to yield interesting lower
bounds in Boolean
depth and, perhaps later, size complexity.

There is a lot of appeal to trying to model Boolean complexity
via cohomology (e.g., the sum of the Betti numbers)
over the appropriate space or topology.
First, cohomology is a natural invariant of spaces for which there is
and wealth of intuition, examples, and tools, some quite sophisticated.  
Second, cohomology often takes infinite or large dimensional
vector spaces and extracts more concise and meaningful information.
Third, cohomology has much overlap with and applications to
combinatorics; toric varieties is one example; more basically,
the inclusion/exclusion principle follows from, via the standard resolution,
the fact that the $n$-simplex (what we below call $\cat{n}$) has the
Betti numbers of a point; so the general study of cohomology (especially
when higher cohomology groups don't vanish) can be viewed as a vast
generalization of inclusion/exclusion.
Fourth, sheaf cohomology and many of its tools makes sense over an
arbitrary topological space or Grothendieck topology, so there
are many possibilities for modelling Boolean functions by sheaves on
Grothendieck topologies.

We emphasize, regarding the fourth point,
that our work here does not involve algebraic geometry
or analysis (e.g., bounds based on degrees and intersection
theory, morse theory, such as used in \cite{steele,benor}).  
Here we will consider only finite topological spaces and, more
generally, Grothendieck topologies whose underlying category is finite
(and ``semitopological'' as defined below).
Such spaces can loosely model some aspects of smooth manifolds while being,
in a sense, not as restricted (or rich in
structure) as manifolds or schemes.  On the other hand, as shown
in [SGA4.I--VI],
any Grothendieck topology
has analogues of sheaves, cohomology, and related concepts
that are strikingly similar to what one is accustomed to from
areas such as analysis, algebraic geometry, group theory, etc.

Let us begin by describing two obstacles to modelling depth complexity
with cohomology.
For any integer $n\ge 1$, consider the Boolean functions on
$n$ variables, $f\from\{0,1\}^n\to\{0,1\}$ (with $1$ being TRUE).
We define the functions $0,1,x_1,\neg x_1,x_2,\ldots,\neg x_n$ to
be of depth\footnote{Morally the functions $0,1$ should probably be defined
also to be of depth $-1$.} $0$, where $x_i$ is the $i$-th coordinate of
$\{0,1\}^n$,
and inductively define a function to be of depth $i$
if it equals $f\wedge g$ or $\neg(f\wedge g)$ with $f,g$ of depth $i-1$
(here $\neg f$ is $1-f$, its logical negation, and $f\wedge g$ is $fg$,
their logical conjunction or AND);
the depth complexity of $f$ is the smallest depth in which it appears.
Let $h$ map
Boolean functions to the non-negative reals, with
\begin{equation}\label{eq:and}
h(f\wedge g) \le c_1+ c_2 \max\bigl( h(f), h(g) \bigr)
\end{equation}
and
\begin{equation}\label{eq:neg}
h(\neg f)=h(f),
\end{equation}
for any Boolean functions $f,g$; then it is easy to see that
$$
\mbox{depth complexity}\ (f) \ge \log_{c_2} \frac{h(f)}{M+\frac{c_1}{c_2-1}},
$$
where $M$ is the maximum of $h(x_1),h(\neg x_1),\ldots,h(\neg x_n)$;
such an $h$ is an example of what is called a {\em formal complexity
measure} (see \cite{wegener}).
Assume that to each Boolean function, $f$, we have associated a topological
space, $U_f$, and we set $h(f)$ to be the sum of the Betti numbers of
$U_f$.  
(Note that we will soon broaden our class of $h$'s, but Betti numbers of
topological spaces is a good place to start.)
We wish to verify equations~(\ref{eq:and}) and (\ref{eq:neg})
for some $c_1,c_2,c_3$.

The first obstacle is that $U,V$ in the plane
can be each diffeomorphic to an open disk, while $U\cap V$ and $U\cup V$
each have an arbitrarily high sum of Betti numbers (make their boundaries
intersect ``wavily'').  This means that if $U_{f\wedge g}$ is
$U_f\cap U_g$ or $U_f\cup U_g$, we do not anticipate that
a general principle
will establish equation~(\ref{eq:and}).  In terms of sheaf theory,
the problem is that the relation between the Betti numbers follows from
the short exact sequence
$$
0\to \myfield_{U_f\cap U_g}\to \myfield_{U_f}\oplus \myfield_{U_g} \to
\myfield_{U_f\cup U_g} \to 0,
$$
where $\myfield_A$ denotes the rationals, $\myfield$, restricted to $A$
and extended by zero elsewhere; since $\myfield_{U_f},\myfield_{U_g}$
control only one (nonzero) term in the sequence, we have no control
on the Betti numbers arising from the other two.
Here we will propose a general principle from which $U_f$ and $U_g$
will control the sequence, providing we pass to Grothendieck topologies
with ``enough virtual zero extensions'' or ``no composition conflicts.''
Once we make these ideas precise, the proof follows immediately from
two short exact sequences.  We shall show that Grothendieck topologies
with ``enough virtual zero extensions'' exist; in fact, so many exist that
we have only begun to study them at this point.  Free categories always
have this property, but they are of homological dimension at most one
and this may indicate that we should look elsewhere for interesting
categories.  Pulling back preserves ``virtual zero extensions'' and can add
new ones, and therefore there are other strategies for building interesting
categories by successive pullbacks, in particular fiber products.

It would be possible that $U_{f\wedge g}$ has ``nothing to do'' with
$U_f$ and $U_g$, so that there really is no obstacle as described above.
But most approaches we have seen in various types of complexity have
some similar relations, and if not
then one still needs a method to establish equation~(\ref{eq:and}).
Note our approach insists nothing about $U_{f\vee g}$, and we deal with
$f\vee g$ indirectly via $f\vee g=\neg((\neg f)\wedge(\neg g))$ (the
minimum depth of a function being closed under negation).

The second obstacle is to establish $h(f)=h(\neg f)$; of course, it
would suffice to establish $h(\neg f)\le c_1 + c_2 h(f)$, but in
most settings $f$ and $\neg f$ have the exact same complexity.  Since
the association between $f$ and $\neg f$ is so natural and simple, we
insist, in this paper, that $h(f)=h(\neg f)$.  Again, it is
possible that the models for $f$ and $\neg f$ have ``nothing to do''
with each other, and that $h(f)=h(\neg f)$ by accident or some less
direct means.  However, we feel that it could be productive to look
for a direct reason for $h(f)=h(\neg f)$.

One possibility is that $f$ and $\neg f$ have the same model.  This does
not work well if for each $f,g$ we have
$U_{f\wedge g}=U_f\cap U_g$, for then
$$
U_f = \bigcap_{\alpha\in\{0,1\}^n,\ \ f(\alpha)=0} U_{\chi_\alpha},
$$
with $\chi_\alpha$ being the characteristic function of $\alpha$
(one at $\alpha$ and zero elsewhere).  In particular, for any
$\beta\ne\gamma$ we have
$$
U_{\chi_\beta}=U_{1-\chi_\beta} = \bigcap_{\alpha\ne\beta}U_{\chi_\alpha}
\subset U_{\chi_\gamma}.
$$
Similarly $U_{\chi_\gamma}\subset U_{\chi_\beta}$, and $U_f$ is
independent of $f$.
Even if we don't have $U_{f\wedge g}=U_f\cap U_g$, it still seems 
hard to deal with equation~(\ref{eq:and}) assuming $f$ and $\neg f$
have the same model.

Another possibility is that $f$ and $\neg f$ could be different but
have the same Betti number sum because of duality between the
cohomology groups of $f$ and $\neg f$, which we will
soon make precise.  Let us mention that this duality is akin to Poincar\'e
or Serre duality.  Furthermore, the main technical result in this
paper is to define what it means for a morphism of the topologies
we study to be ``strongly $n$-dimensional'' (this looks like a
special type of relative
Poincar\'e duality), and to prove that this property
is stable under base change.  This means that the pulling back
that we were considering to obtain ``enough virtual zero extensions''
(to satisfy equation~(\ref{eq:and})) preserves strong dimensionality.

Henceforth we give a more precise description of our approach.


If $X$ is a Grothendieck topology
(for example,
a topological space), and $F,G$ are sheaves of $\myfield$-vector
spaces on $X$, we define
the {\em cohomological complexity} of $(X,F,G)$ as
$$
{\rm cc}_X(F,G)= \sum_{i\ge 0} {\rm dim}_\myfield {\rm Ext}^i_X(F,G).
$$
(In particular, ${\rm cc}_X(\myfield,\myfield)$ is the sum of the 
Betti numbers of $X$.)
By a {\em sheaf model (on $n$ variables)} we mean
an association to each Boolean function on $n$ variables,
$f\from\{0,1\}^n\to\{0,1\}$, a tuple $(X_f,F_f,G_f)$; by the
homological complexity of $f$ we mean that of the associated tuple.



In this paper we limit our focus as follows.
We will consider only the {\em depth complexity} of a Boolean
function.
Furthermore we 
consider only Grothendieck topologies whose underlying category,
$\cc$, is {\em finite}, meaning having finitely many morphisms, and
{\em semitopological}, meaning that the only morphism
from an object to itself is the identity morphism;
in case any two objects have at most one morphism between them,
we shall call $\cc$ {\em of topological type\footnote{Note that the term
``topological category'' has another meaning, namely as a category whose
sets of objects and morphisms are topological spaces with source and target
maps being continuous maps.}}, as the associated topos is
equivalent to one of a topological space, using Theorem~\ref{th:gross}.
Moreover we consider only the {\em grossi\`ere} topology on the category,
$\cc$ (in which the sheaves are just the presheaves), 
since by Theorem~\ref{th:gross} this essentially loses no generality.
The sheaves we will consider will be sheaves of finite dimensional vector
spaces over the rationals, $\myfield$, denoted $\myfield(\cc)$.
The standard resolution
(see Section~\ref{sb:standard_resolution})
then implies that the cohomological complexity is always finite 
in the above case (i.e., the case of
finite dimensional sheaves of $\myfield$-vector spaces
on finite semitopological categories).


Let us return to our first obstacle.
Namely, fix a Grothendieck topology,
$X$, and sheaves $F,G$ for which ${\rm cc}(F,G)=0$.  Then if $U$ is an
open set and $Z$ its closed complement, the short exact sequence
$$
0\to G_U\to G\to G_Z\to 0
$$ 
shows that ${\rm cc}(F,G_U)={\rm cc}(F,G_Z)$.
Consider a sheaf model $f\mapsto (X,F,G_{U_f})$ where $f\mapsto U_f$
is an association of an open set, $U_f$, to each Boolean function, $f$,
with the property that $U_{f\wedge g}=U_f\cap U_g$ for all $f,g$.  
Then the homological
complexity of $f\wedge g$ is bounded by that of $f$ and $g$ provided that
there exists an exact sequence
$$
0\to G_{U_f}\to G_{U_f\cap Z_g}\oplus H\to G_{Z_g}\to 0
$$
for some sheaf, $H$, where $Z_g$ is the closed complement of $U_g$
and where $G_{U_f\cap Z_g}=G\otimes \myfield_{U_f}\otimes \myfield_{Z_f}$
with $\myfield_{U_f},\myfield_{Z_g}$ being the usual open and closed 
extensions by zero.  
Such an
$H$ will exist only for very special $U_f,U_g$ when $\cc$ is topological.
However, such $H$'s exist
whenever $\cc$ has ``enough arrows to avoid composition conflicts''
(see Section~\ref{se:zero_extensions}).

Duality gives a non-trivial equality of cohomological
complexity, given by Ext duality, akin to Serre (or Poincar\'e) duality.
This describes situations where
\begin{equation}\label{eq:hc_invariant}
{\rm cc}(X,F,G) = {\rm cc}(X,G,F')
\end{equation}
for a natural sheaf $F'$ (the star denoting the dual vector space), 
so that if $F'=F\otimes \omega$ for a vector
bundle $\omega$ (see Section~\ref{sb:vector_bundles}), then
$$
{\rm cc}(X,F,G) = {\rm cc}(X,F,G'\otimes\omega^\vee)
$$
where $\omega^\vee$ is the dual sheaf (see Section~\ref{sb:vector_bundles}),
provided that $G'$ exists as well.
More precisely, we shall define a simple functor $!\to *$ on the
derived category $\cd=\cd^{\rm b}(\myfield(\cc))$ (of bounded complexes in
$\myfield(\cc)$)
such that
\begin{equation}\label{eq:shriek_to_star_def}
\Hom_\cd(F,G)= \bigl( \Hom_\cd(G,F^{!\to *}) \bigr)^*
\end{equation}
(we write $F^{!\to *}$ for $(!\to *)F$ at times) for all $F,G\in\cd$
(in other words, $!\to *$ is the Serre functor,
see \cite{bondal_kapranov,bondal_orlov,larsen} and 
Section~\ref{sb:serre}).  When $F,G$ are sheaves
and when $F^{!\to *}\isom F'[n]$ for some $F'$ and $n$, then
equation~(\ref{eq:hc_invariant}) holds.
If $G_0=G_{U_f}$, then we would hope to show that either $G_{U_{\neg f}}$
or $G_{Z_{\neg f}}$ is $G_1=G'_0\otimes\omega^\vee$, or
$G_2=G'_1\otimes\omega^\vee$, etc. (or the same with $G_0=G_{Z_{f}}$).

Let us specialize our discussion.  Consider $F=\myfield$, and consider
those categories $\cc$ for which $\myfield^{!\to *}\isom\myfield[n]$ for 
some $n$.
In this case we say that $\cc$ is {\em strongly $n$-dimensional};
(Some (but not all) categories arising from coverings of manifolds have this
property.)  We can use the product and fiber product to construct new
such categories out of old ones, but our fiber product constructions
sometimes
require a relative notion of strong dimensionality.  Namely, we say that
a functor $f\from\cx\to\cs$ is strongly $n$-dimensional if
\begin{equation}\label{eq:strong_dim_def}
(!\to *)_\cx f^* \isom f^*(!\to *)_\cs[n]
\end{equation}
(if $\cs$ is a point, then it suffices to test this condition on $\myfield$,
which amounts to $\cx$ being strongly $n$-dimensional).
The most difficult theorem in this paper is to show that strong
dimensionality of a morphism is closed under base change (thus giving
a fiber product construction).  This is proven by giving an equivalent,
fiberwise $n$-dimensionality, which is clearly
closed under base change due to its local nature.

It follows that our proposed approaches to equations~(\ref{eq:and}) and
(\ref{eq:neg}) are both ``compatible'' with pulling back or fiber
products in some sense.  We hope that fiber products and pull backs,
combined with a sufficient collection of examples (as we begin to
establish in Section~3) will yield interesting sheaf models.

Before describing the rest of this paper, we make
two remarks.  First, the size of the categories needed
to have enough ``virtual zero extensions'' seems to be (in number of
morphisms) doubly or triply exponential in $n$, meaning that our
techniques are not ``natural'' in the sense of \cite{razborov}.
Finally, Mulmuley and Sohoni have an approach to circuit complexity
that is very different from ours in that it uses algebraic geometry,
in particular geometric invariant theory; see the series of papers
beginning with \cite{mulmuley}.

In Section~2 we fix a lot of notation and recall various facts needed
later; all facts are known or follow easily from known results.
In Section~3 we describe how certain topological spaces (e.g., smooth
manifolds) are ``modelled''
by categories in that their Betti numbers agree; our modelling can
give rise to categories that are not of topological type, and rather
than just involving open covers, our modelling
also involve {\espaces}, i.e., local
homeomorphisms, which accounts for the fact that there can be more than
one arrow between two objects in the associated category.
Section~4 discusses virtual zero extensions in more detail and their
relationship to equation~(\ref{eq:and}).  Section~5 classifies the
injectives and projectives of $\myfield(\cc)$ for $\cc$ finite and
semitopological, from which the functors $!\to *$ and $*\to !$ are
defined.  Section~6 proves that $!\to *$ is the Serre or duality
functor; we note that this result is similar to duality theory
for toric varieties (see \cite{barthel} and the references there).  
Section~7 gives a necessary (but not sufficient) linear
algebraic condition for $G=(!\to *)F$ to hold, based on the ``local
Euler characteristics'' of $F$ and $G$.
Section~8 states the theorem that strong $n$-dimensionality, a
compatibility condition of $!\to *$ with pulling back, is stable
under arbitrary base change; fiberwise $n$-dimensionality is also
defined and is shown to be stable under arbitrary base change;
Section~9 proves that strong dimensionality is equivalent to
fiberwise dimensionality.
Section~10 investigates the base change morphisms, related to
those used in Section~9.  Appendix~A formulates duality abstractly,
in hopes that we might find other interesting dualities and to
put the duality used in this paper on a firmer foundation.

We wish to acknowledge a number of people for discussions; on the
literature: Kai Behrend, Jim Bryan, Jim Carroll, Bernard Chazelle,
Sadok Kallel,
Kalle Karu, Kee Lam, Laura Scull, Janos Simon, and Steve Smale;
on the exposition: Lenore Blum, Avner Friedman, Richard Lipton,
and Satya Lokam;
and finally Denis Sjerve whose example with ``multiple wrapping'' around the
circle lead to the example at the end of Section~3.

\section{Preliminary Remarks and Notation}

In this section we make some preliminary remarks regarding this paper
that are either known or easy, and we fix our notation.

If $\cc$ is a category, then $\objects{\cc}$ denotes the objects of 
$\cc$ and $\morphisms{\cc}$ denotes the morphisms of $\cc$.
If $\phi\in\morphisms{\cc}$, then $s\phi$ denotes the source of $\phi$,
and $t\phi$ denotes the target.

\subsection{Adjoints to the Pullback}
\label{sb:adjoint}
A finite or infinite sequence $\ldots,u_0,u_1,\ldots$ of functors is said
to be a {\em sequence of adjoints} if we have
$u_i$ is the left adjoint of $u_{i+1}$ for all relevant $i$.

If $\cc$ is a category, then $\myfield(\cc)$ (respectively, $\widehat\cc$)
denotes the category of
presheaves on $\cc$ with values in (i.e., the category of functors 
from $\cc^{\rm opp}$ to) the category of finite dimensional
$\myfield$-vector spaces (respectively, the category of sets).
If $u\from \cc\to\cc'$ is a functor between finite categories, and
$u^*\from \myfield(\cc')\to\myfield(\cc)$ is the pullback, then
according to [SGA4.I.5.1], $u^*$ has a left adjoint $u_!$ and a right
adjoint $u_*$.  We shall denote by $u^?$ (respectively, $u^!$) the
left adjoint to $u_!$ (respectively, right adjoint to $u_*$) when
they exist.  [SGA4.I.5.6] shows that any of $u,u_*,u_!$ being fully
faithful implies that the other two are, and that this condition is
equivalent to either adjuction morphism ${\rm id}\to u^*u_!$ or
$u^*u_*\to {\rm id}$ being an isomorphism.

Let us spell
out $u^*,u_*$, the adjoint mappings,
and the adjunctive morphism $\id\to u_*u^*$.  Let $F\in\myfield(\cc')$
and $G\in\myfield(\cc)$.
For $X\in\objects{\cc}$ we have $(u^*F)(X)=F(u(X))$, and for 
$Y\in\objects{\cc'}$ we have
$$
(u_*G)(Y)= \lim_{\substack{{\longleftarrow}\\{X;u(X)\to Y}}} G(X),
$$
where the limit is over the category whose objects are pairs $(X,m)$
with $m\from u(X)\to Y$ (see [SGA4.I.5.1]).  Next we describe
$$
\mu\from\Hom(u^*F,G)\to\Hom(F,u_*G);
$$
if $\phi\in\Hom(u^*F,G)$, then we have maps
$$
\phi_X\from (u^*F)(X)=F(u(X))\to G(X),
$$
and the map
$$
(\mu\phi)_Y\from F(Y)\to 
\lim_{\substack{{\longleftarrow}\\{X;u(X)\to Y}}} G(X)
$$
is simply given by
$$
F(Y)\to \lim_{\substack{{\longleftarrow}\\{X;u(X)\to Y}}}F(u(X))
\ \ \to \lim_{\substack{{\longleftarrow}\\{X;u(X)\to Y}}}G(X),
$$
where the first arrow is uniquely determined from the definition of limit,
and the second arrow arises from applying $\phi_X$ to each $F(u(X))$.
The quasi-inverse to $\mu$, $\nu$, is given on $\psi\in\Hom(F,u_*G)$
via
$$
F(u(X))\ \ \to\ \  \lim_{\substack{{\longleftarrow}\\{Z;u(Z)\to u(X)}}}G(Z)
\ \ \to\ \  G(X),
$$
where the first arrow is given by $\psi_{u(X)}$, and the second by
the canonical map of the limit onto $G(X)$  corresponding to the
object $(X,\id_X)$ (in the category over which the limit is taken).
See [SGA4.I.5.1] for details of the above (there they discuss only
$u_!$, where the arrows are reversed).

Setting $G=u^*F$, it follows that the adjuction morphism $\id\to u_*u^*$
is given by the natural map
$$
F(X)\to \lim_{\substack{{\longleftarrow}\\{Z;u(Z)\to X}}}F(u(Z)).
$$

In the above, we have implicitly
touched on a number of properties of limits.  Another fact we will use
is that if in addition we have $v\from\cc'\to\cc''$ with $\cc''$ finite,
then there is a canonical isomorphism
$$
\lim_{\substack{{\longleftarrow}\\{Y;v(Y)\to Z}}}\ \ \ 
\lim_{\substack{{\longleftarrow}\\{X;u(X)\to Y}}} G(X) \isom
\lim_{\substack{{\longleftarrow}\\{X;(vu)(X)\to Z}}} G(X)
$$
(this can be verified directly, or follows because
$(vu)_*$ is canonically isomorphic to $v_* u_*$, using Yoneda's
lemma and that $(vu)^*=u^*v^*$).

If $P\in\objects{\cc}$, then $k_P\from\cat{0}\to\cc$ denotes the map
from the one object, one morphism category, $\cat{0}$, to $\cc$ sending
the object of $\cat{0}$ to $P$.  
For a $\myfield$-vector space, $V$,
[SGA4.I.5.1] shows that $k_{P!}V$
is isomorphic to the sheaf whose value at $Q\in\objects{\cc}$ is
\begin{equation}\label{eq:nitpicky}
(k_{P!}V)(Q)=V^{\Hom_\cc(Q,P)};
\end{equation} 
notice that although [SGA4.I.5.1] defines $k_{P!}$ as a limit, and therefore
ambiguous up to isomorphism, we shall chose $k_{P!}$ to mean equality
in equation~(\ref{eq:nitpicky}) (this may seem nitpicky, but it will be
necessary to chose one version of $k_{P!}$ to define $!\to *$ in
Section~\ref{se:injectives});
a morphism $\eta\from Q_1\to Q_2$ 
gives a map from $\Hom(Q_2,P)$ to $\Hom(Q_1,P)$,
giving rise to a morphism
$$
V^{\Hom(Q_1,P)} \to V^{\Hom(Q_2,P)}
$$
and its transpose
$$
(k_{P!}V)(\eta)\from (k_{P!}V)(Q_2)\to (k_{P!}V)(Q_1).
$$
The functor $k_{P*}$ is the same with arrows reversed,
e.g., replace $\Hom_\cc(Q,P)$ with 
$$\Hom_\cc(P,Q)=\Hom_{\cc^{\rm opp}}(Q,P)
$$
(but the map $(k_{P*}V)(\eta)$ is defined directly, without the 
transpose).

It will be important to study how adjoint functors give rise to adjoints
in the derived categories.  Let a functor $u\from\ca\to\cb$ have right adjoint 
$v\from\cb\to\ca$, where $\ca,\cb$ are Abelian categories.
Let $A^\bullet$ be a complex in $\ca$ and $B^\bullet$ one in $\cb$.
By $uA^\bullet$ we mean the complex whose $i$-th element is $uA^i$, and
similarly for $vB^\bullet$.
A morphism of complexes $uA^\bullet\to B^\bullet$ gives arrows
$uA^i\to B^i$, that in turn give maps $A^i\to vB^i$; it is easy to check
that these maps give a morphism of complexes $A^\bullet\to vB^\bullet$
that preserves homotopies.  We can invert this procedure, and therefore
conclude that $u,v$ are adjoints in $\ck(\ca),\ck(\cb)$ (the categories of
complexes with morphisms being chain maps modulo homotopy),
i.e., we have a bi-natural isomorphism in the variables $A^\bullet,B^\bullet$
$$
\Hom_{\ck(\cb)}(uA^\bullet ,B^\bullet) \isom
\Hom_{\ck(\ca)}(A^\bullet ,vB^\bullet).
$$
If either $uA^\bullet$ is a complex of injectives or $B^\bullet$ is a
complex of projectives, then
$$
\Hom_{\ck(\cb)}(uA^\bullet ,B^\bullet) = \Hom_{\cd}(uA^\bullet,B^\bullet),
$$
where $\cd$ is any of $\cd(\cb),\cd^+(\cb),\cd^-(\cb),\cd^{\rm b}(\cb)$
as appropriate.  A similar remarks holds for $\Hom(A^\bullet,vB^\bullet)$.
We conclude (among other similar remarks)
that if any
element of $\ca$ or $\cb$
has a bounded injective resolution and a bounded projective resolution
then $\lder u,\rder v$ are adjoints in $\cd^{\rm b}(\ca),\cd^{\rm b}(\cb)$.

Here is another remark on adjoints that we shall use.
Let $u\from\ca\to\cb$ be a fully faithful functor with right adjoint, $v$,
Then we claim that the adjunctive map $\id\to vu$ is an isomorphism
(as mentioned in the proof of [SGA4.I.5.6]).  Indeed this follows from
Yoneda's lemma and the
bi-natural isomorphism in $A_1,A_2$
$$
\Hom_{\ca}(A_1,A_2) \isom \Hom_\cb(uA_1,uA_2) \isom \Hom_\ca(A_1,vuA_2).
$$
More generally, let $\tilde u\from\tilde\ca\to\cb$ have right adjoint, $v$,
such that the image of $v$ is contained in the subcategory $\ca$ of
$\tilde\ca$, and with $u=\tilde u|_\ca\from\ca\to\cb$ fully faithful.
We claim that the adjuctive map $\id\to v\tilde u$ restricted to
$\ca$ is an isomorphism
on each object of $\ca$, and is the same as $\id\to vu$.  Indeed
for $A\in\objects{\ca}$ we have
$$
\Hom_{\cb}(uA,B) = \Hom_{\cb}(\tilde uA,B) \isom \Hom_{\tilde\ca}(A,vB)
= \Hom_{\ca}(A,vB),
$$
which shows that $v$ is also a right adjoint to $u$ with the bi-natural
isomorphism $\Hom_\cb(uA,B)\to\Hom_\ca(A,vB)$ of the $u,v$ adjointness
being the restriction that of the $\tilde u,v$ adjointness.
Hence the adjunctive map $A\to vuA$, which is the image of $\id_{uA}$,
is the same as the adjunctive map $A\to \tilde vuA$.  Finally from the
above we know that $A\to vuA$ is an isomorphism.

\subsection{Partial Order and Primes}

Consider a semitopological category, $\cc$ (as in the introduction, this
means that any morphism from an object to itself is the identity morphism of
that object).  For $U,V\in\objects{\cc}$ we write $U\le V$ or $V\ge U$
if there is
a morphism from $U$ to $V$.  This is a semi-partial order, meaning that
it is a partial order except for that we may have $U\le V$ and $V\le U$
without $U$ and $V$ being the same object (but then $U$ and $V$ must be
isomorphic).  If the category is {\em sober},
meaning that any two isomorphic objects are equal, then the semi-partial
order becomes a partial order.

Throughout this paper, when we speak of ``greater,'' ``increasing chains,''
etc., we mean with respect to this semi-partial order.  

To {\em factor} a morphism means to write it as a composition of two
or more morphisms.  A {\em prime} is a nonidentity morphism that cannot
be factored into two nonidentity morphisms.  A functor is determined
by its action on the objects and prime morphisms, assuming every morphism
can be factored into a finite number of primes.

\subsection{Composable Morphisms}
\label{sb:composable}
In a category, $\cc$, we use 
$\fleches{0}{\cc}$ to denote $\objects{\cc}$.  For any integer
$i\ge 1$, we use $\fleches{i}{\cc}$ to denote the set of $i$-tuples
of morphisms $(\phi_1,\ldots,\phi_i)$ that are {\em composable},
meaning that $s\phi_k=t\phi_{k-1}$ for $k=2,\ldots,i$ (so that
$\phi_i\circ\cdots\circ\phi_1$ exists); in particular 
$\fleches{1}{\cc}=\morphisms{\cc}$.

For integer $m\ge 0$, let $\cat{m}$ denote the category whose objects
are $\{0,\ldots,m\}$ and with one or zero (respectively) 
morphisms from $i$ to $j$ according to whether or not $i\le j$.
We often call $\cat{m}$ the {\em $m$-dimensional simplex}.
A functor, $F$, from $\cat{m}$ to a category $\cc$ is determined by the
$m$ composable morphisms $F(0\to 1),\ldots,F(m-1\to m)$.
Let $\underfleches{m}{\cc}$ be the category whose objects are functors
from $\cat{m}$ to $\cc$ (and whose morphisms are natural transformation);
clearly the objects of $\underfleches{m}{\cc}$ can be identified with
$\fleches{m}{\cc}$.

We extend the definition of $\cat{m}$ and
$\fleches{m}{\cc}$ to $m=-1$ by setting
$\cat{-1}$ to be the empty category, making $\underfleches{-1}{\cc}$ to
be a {\em ponctuel} category of one object and one morphism.

For integer $m\ge -1$ we define the usual $m+2$ functors
${\rm coface}_i\from \cat{m}\to\cat{m+1}$ determined, for $m\ge 0$
and $i=0,\ldots,m+1$,
by the map on
objects (since $\cat{m+1}$ is a partial order),
$$
{\rm coface}_i(j) = \left\{\begin{array}{ll} j & \mbox{if $j<i$,} \\
j+1 & \mbox{otherwise.} \end{array}\right.
$$
Then ${\rm face}_i = {\rm coface}_i^*$ gives rise to the usual simplicial
complex
$$
\fleches{-1}{\cc}\leftarrow\fleches{0}{\cc}
\twoleftarrows\fleches{1}{\cc}
\threeleftarrows\fleches{2}{\cc}\cdots,
$$
where the faces of $(\phi_1,\ldots,\phi_m)\in \fleches{m}{\cc}$
(in order, from ${\rm face}_0$ to ${\rm face}_m$,) are
$$
(\phi_2,\ldots,\phi_m), \; (\phi_2\circ\phi_1,\phi_3,\ldots,\phi_m),\;
(\phi_1,\phi_3\circ\phi_2,\phi_4,\ldots), \; \ldots ,
$$
$$
(\phi_1,\ldots,\phi_{m-2},\phi_m\circ\phi_{m-1}),\;
(\phi_1,\ldots,\phi_{m-1})
$$
(see, for example, [SGA4.V.2.3.6]).

\subsection{Simiplicial Complex, Simplicial Hom, and Graphs}
\label{sb:simplicial}

In the previous subsection we have described a map taking
a category and returning a simplicial complex; denote this map $u$.  
Furthermore, there
is a map, $v$, that associates to each simplicial complex its
associated graph, by forgetting about the sets of dimension two
and greater.  The map $v\circ u$ is the usual forgetful functor
from categories to graphs; its left adjoint is the free category
associated to a graph, which associates to a graph the category
whose morphisms are walks in the graph (and objects being the vertices
of the graph).

For example, the free category associated to a directed path of
length $n$ is the category $\cat{n}$.
A category is isomorphic to a free category
iff it is a {\em unique factorization domain},
i.e., each nonidentity morphism can be factored as a finite number
of primes in exactly one way.

Given a category, $\cc$, and an additive category, $\cd$, we define
a category $\simexp{\cc}{\cd}$, the {\em simplicial Hom of $\cc$ to
$\cd$} as follows: its objects are the set theoretical maps from
$\objects{\cc}$ to $\objects{\cd}$; given objects $F,G$, an element
of $\Hom(F,G)$ is a map $\alpha\from\morphisms{\cc}\to\morphisms{\cd}$
such that $s\circ\alpha=F\circ s$ and same with $t$ replacing $s$ and
$G$ replacing $F$; in
other words, for each morphism $\phi\in\morphisms{\cc}$, $\alpha\phi$
is an arrow $Fs\phi\to Gt\phi$.
The composition is defined via
$$
(\beta\alpha)(\phi)=\sum_{\phi_2\phi_1=\phi} (\beta\phi_2)(\alpha\phi_1).
$$
The morphisms and their compositions can be viewed as a generalization
of matrices with
matrix multiplication over $\cd$ indexed in the objects of $\cc$.
For an object, $F$, we have $\id_F$ is the map $\id_F(\id_X)=\id_{F(X)}$
and $\id_F(\phi)=0$ if $s\phi\ne t\phi$.

\subsection{The Grossi\`ere Topology}

In this subsection we show that the category of sheaves of sets of any
finite semitopological Grothendieck topology, $(\cc,J)$, 
is equivalent to the category of presheaves of sets (on a certain full
subcategory of $\cc$).

Let $E=({\cal C},J)$ be a Grothendieck topology or site.
We say that $X\in\objects{\cal C}$ is
{\em gross} if $J(X)=\{ X\}$ (i.e., the only element of $J(X)$
is the sieve that is the entirety of ${\cal C}/X$).
Recall that the {\em grossi\`ere} (meaning gross or coarse) topology
([SGA4.II.1.1.4])
for a
category $\cc$ is the Grothendieck topology for which each object is
gross; 
in this topology a sheaf is the same thing as a presheaf (as defined
here and by Grothendieck, [SGA4.I.1.2]).

If $U$ is an open set in a topological space, then $U$ is
gross iff it
is irreducible, i.e., iff $U$ is not the union of its proper open subsets,
iff there is a point, $p\in U$, such that any open set containing $p$
contains $U$.

\begin{theorem}\label{th:gross}
Let $E=(\cc',J')$ be a finite Grothendieck topology.
The gross objects of $\cc'$ determine a full subcategory, $\cc$;
let $u\from\cc\to\cc'$ be the inclusion.
Then $u^*$ gives an
equivalence of categories between $\widehat\cc$ and
the category of sheaves on $E$.
\end{theorem}

\proof Let $J$ be the topology induced by $u$ on $\cc$.  First we show
that $J$ is the {\em grossi\`ere} topology.  Let $R\in J(Y)$.
Since $u$ is fully faithful we have $u^*u_!R=R$; it follows that
$(u_!R)(Y)=R(Y)$.  But the image of $u_!R\to Y$ is {\em bicouvrant}
by [SGA4.III.3.2],
and in particular $(u_!R)(Y)=\{ \id_Y \}$.  It follows that $R(Y)=\{ \id_Y\}$,
and so $R=Y$.  Hence $J(Y)=\{ Y\}$, and $J$ is the {\em grossi\`ere}
topology.

We finish by showing that $u$ satisfies condition (i) of
the hypothesis of the Comparison Lemma
([SGA4.III.4.1]); condition (ii) then follows, which is the claim
of our theorem.  If $\{ H_i\to K\}$ is a {\em couvrante} family,
and $L_i\to H_i$ is {\em couvrant} for each $i$, then $\{ L_i\to K\}$
is a {\em couvrante} family, by [SGA4.II.5.1.ii].  
This remark allows us to take any {\em couvrante} family, $\{ Y_i\to X\}$,
where the $Y_i$ are objects, take any $Y_k$ that is not gross
and maximal among the non-gross, and replace it with objects
less than it (here ``maximal'' and ``less'' refer to the partial order
$V\le W$ if there is a morphism from $V$ to $W$).
It follows (since $\cc'$ is finite)
that any object in $\cc'$ can be covered\footnote{
The word {\em recouvrement}, used in [SGA4.II.5.1], does not appear to be
defined up to that point; however, it is clear from the proof
(especially where $pn=qn$ implies that the kernel of $p,q$ is a
{\em couvrant} sieve) (and from [SGA4.V.2.4.3]) that
a set of objects,
$\cs$, is a {\em recouvrement} of $X$
if there is a family of morphisms with sources in $\cs$
and target $X$ that is a {\em couvrante} family for $X$.
(Also, there is no word {\em couvrement} in French, so in making a noun
out of {\em couvrir} one must choose between {\em recouvrement} and
{\em couverture}, the latter not sounding very mathematical.)
} 
by morphisms from
gross objects.
\proofbox

\subsection{Topological Notions}

The following notions agree with [SGA4.IV] in the case where a category
is endowed with the {\em grossi\`ere} topology, which is our running
assumption for most of this paper.  A point in a
sober, finite, semitopological category, $\cc$, is an object of $\cc$
(see [SGA4.IV.6.8.6]).  An {\em open set} of $\cc$ (see
[SGA.IV.8.4.4]) is a 
{\em sieve}, i.e., a full
subcategory, $\cc'$, of $\cc$, such that if $f\in\morphisms{\cc}$ and
$tf\in\objects{\cc'}$, then $sf\in\objects{\cc'}$; a sieve is determined
by its objects, and we sometimes identify the sieve with its set of
objects (if no confusion will arise).  {\em Closed sets} and {\em cosieves}
are defined dually, i.e., as open sets and sieves (respectively) in
$\cc^{\rm opp}$.
The {\em complement} of a subcategory, $\cc$, of $\cc'$ is the full
subcategory of $\cc'$ whose objects are $\objects{\cc'}\setminus
\objects{\cc}$. 

If $j\from U\to X$ is the inclusion of an open subcategory, $U$, of a
category, $X$, and $G\in\myfield(U)$, then $j_!G$ is the usual extension
by $0$ of $G$; if $F\in\myfield(X)$, then the left adjoint, $j^?$, of
$j_!$ satisfies
$$
(j^?F)(V) \isom
\{ \ell\in F(V)^* \;|\; \ell(F\phi)=0\quad\forall \phi\from V\to A,\ 
{\rm with}\ A\notin U
\}^* 
$$
$$
\isom F(V) \biggm/ \bigoplus_{\phi\from V\to A,\ 
{\rm with}\ A\notin U} {\rm im}(F\phi)\ \ 
\isom {\rm coker}\ \  \bigoplus_{\phi\from V\to A,\ 
{\rm with}\ A\notin U} F\phi
$$
Similarly if $i\from Z\to X$ is a closed inclusion, then $i_*$ is the
usual extension by $0$, and has right adjoint, $i^!$ with
$$
(i^!F)(V)\ \  \isom\ \ 
\ker \bigoplus_{\phi\from A\to V,\ {\rm with}\ A\notin Z} F\phi
$$
(often called ``sections supported on $Z$'').

\subsection{Simple Duality}

For a presheaf, $F$, of finite dimensional
$\myfield$ vector spaces on a category, ${\cal C}$,
we define the presheaf of ${\cal C}^{\rm opp}$, $F^{\rm dl}$ as follows:
first,
$$
F^{\rm dl}(U) = {\rm Hom}_{\myfield}(F(U),\myfield);
$$
second, a morphism, $\phi\from U\to V$ in ${\cal C}$ corresponds to
a morphism $\phi^{\rm opp}\from V\to U$ in ${\cal C}^{\rm opp}$,
and we define $F^{\rm dl}\phi^{\rm opp}$ to be the map dual to $F\phi$.

\begin{theorem} The functor ``dual'' is exact, involutive, and exchanges
injectives for projectives and vice versa.
Furthermore, for a full inclusion of 
categories, $k$, we have $(k_*)\circ\,{\rm dl}=\,{\rm dl}\circ k_!$.
\end{theorem}

By passing to the ``dual'' of a sheaf on the opposite category, we
can often prove two theorems at once.

\subsection{Vector Bundles}
\label{sb:vector_bundles}

By a {\em vector bundle} we mean an $F\in\myfield(\cc)$ such that
$F\phi$ is an isomorphism for all $\phi\in\morphisms{\cc}$.
Associated to $F$ is its {\em dual bundle}, $F^\vee$, such that
for $X\in\objects{\cc}$ we have
$F^\vee(X)$ is the dual space to $F(X)$, and for $\phi\in\morphisms{\cc}$
we have $F^\vee\phi$ is the inverse of the dual to $F\phi$.
A {\em line bundle} or {\em invertible sheaf} is a line bundle for which
$F(X)\isom\myfield$ for each $X\in\objects{\cc}$.

This notion of vector bundle is justified as follows.
If $X\in\objects{\cc}$, then by {\em localization at $X$}
we mean the ``source'' map $j_X\from \cc/X\to \cc$ 
([SGA4.I.5.10--12,III.5,IV.8]).
(So if $\cc$ is topological, then $\cc/X$ can be identified with
the open subset associated with $X$, and $j_X$ is the usual inclusion
of categories.)
By a {\em vector bundle} we mean an $F\in\myfield(\cc)$ that is locally
trivial, i.e., such that
for each $X\in\objects{\cc}$ we have $j_X^*F$ is isomorphic to a number
of copies of $\myfield$; it is easy to see that this is equivalent to
the definition of vector bundle in the previous paragraph.

If $F$ is a vector bundle, then there is a {\em dual vector bundle}, $F^\vee$,
given by
$$
X\mapsto F^\vee(X)=\Hom( j_X^*F,\myfield) ,
$$
and for $\phi\from X\to Y$ in $\cc$ we determine $F^\vee\phi$ by the
functor $\cc/Y\to\cc/X$ arising from $\phi$.
This definition is equivalent to the previous one.

We remark that for any vector bundle, $F$, in $\cc$, and $G,H\in\myfield(\cc)$,
we have
$$
{\rm Ext}^i(F\otimes G,H) \isom {\rm Ext}^i(G,H\otimes F^\vee),
$$
since both left- and right-hand-sides are delta functors in $H$ that are
isomorphic for $i=0$.
(More generally, if $A^\bullet,B^\bullet$ are bounded complexes, then
$\Hom(F\otimes A^\bullet,B^\bullet)\isom\Hom(A^\bullet,B^\bullet\otimes
F^\vee)$ in the category, $\ck^{\rm b}(\myfield(\cc))$, of bounded
complexes over $\myfield(\cc)$, whose morphisms are chain maps modulo
homotopy.)

\subsection{Abstract Principles}

\subsubsection{Usage of the Axiom of Choice}
\label{sb:axiom of choice}
Some functors constructed in this paper, notably some quasi-inverses
and Serre functors, have freedom in their definition and require
choices to be made definite.  At first glance it seems we require the
Axiom of Choice 
(e.g.,
Axiom~(\underline{U}B) of [SGA4.I.1], page 3)
for this.  However, in practice we are interested in the
behavior of these functors only on a finite number of objects (and morphisms
between these objects).  It is not hard to see that it suffices to
apply the Axiom of Choice to subcategories with a finite number of
objects, whereby the axiom of chioce is not needed.  Let us give an
example.

Say that $F\from\cx\to\cy$ is fully faithful and essentially surjective,
and say that we wish to construct a quasi-inverse, $G$, to $F$, without
invoking the Axiom of Choice.  Say that we are interested to know
$G$ on only a 
subcategory of $\cy$, $\cy'$, that has only finitely many objects.
Then it is easy to see that we can find subcategories $\cx'',\cy''$
(respectively) of $\cx,\cy$ (respectively) such that (1) each has
finitely many objects, (2) $\cy'$ is a subcategory of $\cy''$,
(3) $F$ maps $\cx''$ to $\cy''$ fully faithfully and essentially surjectively.
The quasi-inverse of $F$ restricted to $\cx'',\cy''$ can be constructed
with only a finite number of choices.

In the last paragraph, one can say that $F$ (as a functor $\cx\to\cy$)
is an {\em extension} of $F|_{\cx''}$ ($F$'s restriction as a functor
$\cx''\to\cy''$), or that $F|_{\cx''}$ is a {\em restriction} of $F$.
Functors are partially ordered with respect to extension.  
The last paragraph makes it look like we need to fix $\cy'$ or $\cy''$
once and for all.  However,
if one sees that one
needs a quasi-inverse to $F$ on a larger subcategory than $\cy'$ or $\cy''$, 
then one is free to
extend the quasi-inverse a finite number of times (to successively larger
categories provided each has a finite number of objects) using only finite
choices.  We won't state a formal result, just note that we need to
extend categories so that we can find
choice data (see Section~\ref{sbsb:representative} below) for
the new objects of $\cy$ on which we want the quasi-inverse defined.


\subsubsection{Equivalence of Compositions of Functors}
If $E\from\cc\to\cc$ is an equivalence of categories, with
$\phi$ the invertible natural transformation from $\id_\cc$ to $E$, and if
$F,G$ are functors such that $FEG$ exists, then $FG\isom FEG$ by
horizontally composing $\id_F\phi\id_G$ on $F\id_\cc G$.
It follows (by vertical composition) that one composition
of functors is isomorphic to another iff the same is true when we insert
arbitrary equivalences of categories into the compositions.

\subsubsection{Representative Subcategories}
\label{sbsb:representative}
We say that a full subcategory, $\cc'$, of a category, $\cc$, is a 
{\em representative subcategory} if every object of $\cc$ is isomorphic
to at least one object of $\cc'$; by {\em choice data, $(Z,\iota)$} 
for such a situation
we mean a map $Z\from\objects{\cc}\to\objects{\cc'}$ 
and $\iota\from\objects{\cc}\to\morphisms{\cc}$ such that for each
$X\in\objects{\cc}$, $\iota(X)$ is an isomorphism from $X$ to $Z(X)$.

Given a representative subcategory, choice data always exists provided
we are willing to invoke the Axiom of Choice (but see 
Section~\ref{sb:axiom of choice}).  Alternatively, sometimes the
data, or at least part of it, can be made explicit.  Finally, we remark
that sometimes we want the choice data, especially the morphisms $\iota(X)$,
to satisfy additional constraints (for example, in our construction
of $!\to *$).

In a number of situations in this paper, notably with the derived 
category, it is much simpler to work with a representative subcategory
(in defining functors and natural transformations).
General principles say that we can extend the functors and natural
transformations to the original category.  Here we carefully state these
general principles, at least those that we use in this paper.

\paragraph{Functor extensions}
Given a representative subcategory, $\cc'$, of $\cc$, with choice data
$(Z,\iota)$, and given a functor $F\from\cc'\to\ce$, we define
$F'\from\cc\to\ce$, called {\em $F$'s extension to $\cc$} via:
$$
F'(X) = F\bigl( Z(X) \bigr)
$$
for $X\in\objects{\cc}$, and for $\phi\from X_1\to X_2$ in $\morphisms{\cc}$
set
$$
F'(\phi) = F\bigl(  \iota(X_2)\circ\phi\circ\iota(X_1)^{-1} \bigr).
$$
This construction is absolutely standard (it is essentially how 
quasi-inverses are constructed).  It is standard and easy that if
$F''$ is an extension formed by other choice data, then $F'\isom F''$.

\paragraph{Natural Transformation Extensions}
\label{pa:nte}
Consider a representative subcategory, $\cc'$, of $\cc$, with choice data
$(Z,\iota)$.
Let $F,G$ be two functors from $\cc$ to $\ce$.  Let $\phi$ be a
natural transformation from $F|_{\cc'}$ (i.e., $F$ restricted to $\cc'$)
to $G|_{\cc'}$.  We can extend $\phi$ to a morphism $\phi'\from F\to G$
by setting
$$
\phi'(X) = \bigl(G\iota(X)\bigr)^{-1}\phi\bigl(Z(X)\bigr)F\iota(X).
$$
We easily verify that $\phi'$ is a natural transformation, since 
for $f\from X\to Y$ in $\cc$, each
of the small squares in the diagram below commute:
$$
\begin{CD}
FX @>{\phi'(X)}>> GX \\
@V{F\iota X}VV @VV{G\iota X}V \\
F(Z(X)) @>{\phi(Z(X))}>> G(Z(X)) \\
@V{F(\iota(Y)f\iota^{-1}(X))}VV   @VV{G(\iota(Y)f\iota^{-1}(X))}V \\
F(Z(Y)) @>{\phi(Z(Y))}>> G(Z(Y)) \\
@V{F\iota^{-1}Y}VV @VV{G\iota^{-1}Y}V \\
FY @>{\phi'(Y)}>> GY
\end{CD}
$$

\subsection{The Standard Resolution}
\label{sb:standard_resolution}

Let $\cc$ be a category.  Let $\cf\subset\morphisms{\cc}$
be such that $\phi\notin\cf$ implies that $\phi=\id_X$ for
some object $X$ such that the only morphism from $X$ to itself is the
identity.  For example, if $\cc$ is semitoplogical, then $\cf$ can be
any collection of morphisms that includes all nonidentity morphisms.
Also, regardless of $\cc$, we can always take $\cf=\morphisms{\cc}$,
which is the usual standard resolution ([SGA4.V.2.3.6]).

Let $\cf^i$ be the composable $i$-tuples of
morphisms.  For $F\in\myfield(\cc)$ set
$$
P_i = P_i(F) =
\bigoplus_{\phi\in\cf^i} k_{s(\phi)!}  F\bigl(t(\phi)
\bigr)
$$
and
$$
I^i = I^i(F) =
\bigoplus_{\phi\in\cf^i} k_{t(\phi)*}  F\bigl(s(\phi)
\bigr).
$$
The structure of the $\cf^i$ as a simplicial set 
(see Subsection~\ref{sb:composable} or [SGA4.V.2.3.6], for example)
gives 
complexes\footnote{Note that morally speaking we are saying that
$P_{-1}$ is $F$.  We are not entirely sure why.
Perhaps, since $P_i$ is a sum of $k_{s!}k_t^*F$,
when there is no $s,t$ the $k$'s should be omitted, leaving just $F$?
We leave this to experts on the empty category, $\cat{-1}$$\ldots$}
$$
\cdots\to P_1\to P_0\to F
$$ 
and 
$$
F\to I^0\to I^1\to\cdots
$$
We claim these complexes
give a projective resolution and an injective resolution respectively.  
Too see this,
by simple duality it suffices to check the case of the $P_i$'s.  To
check the exactness at $Y\in\objects{\cc}$, it suffices to find a chain
homotopy $K$ such that $dK+Kd={\rm id}$.  Now
$$
P_i(Y) = \bigoplus_{\phi\in\cf^i} F\bigl(t(\phi) \bigr)^{\Hom\bigl( Y,
s(\phi)\bigr)} = 
\bigoplus_{\mu;\phi_1,\ldots,\phi_i} F\bigl( t(\phi_i) \bigr),
$$
where the rightmost direct sum ranges over $\mu\in\morphisms{\cc}$ and
$\phi_1,\ldots,\phi_i\in\cf$ such that $\mu,\phi_1,\ldots,\phi_i$ is
composable.  If $w\in P_i(Y)$ we define $Kw$ via its components as
$$
(Kw)_{\alpha;\mu,\phi_1,\ldots,\phi_i} = 
\left\{ \begin{array}{ll} w_{\mu;\phi_1,\ldots,\phi_i} & \mbox{if $\alpha=
{\rm id}_Y,$} \\ 0 & \mbox{otherwise.} \end{array} \right.
$$
If $w\in F(Y)$ we define $(Kw)_{{\rm id}_Y;}=w$ and define $Kw$ to vanish
on all other components.  We easily verify $dK+Kd={\rm id}$ on the
complex in question.

In particular, let $\cc$ be a semitopological category, and let
the {\em dimension} of $\cc$, denoted $d=d(\cc)$, be the length
of the longest sequence of composable nonidentity morphisms in $\cc$.
Alternatively, $d+1$ is the length of the longest chain in
the partially ordered set of objects.  Alternatively $d$ is the
{\em topological dimension}, meaning that $d+1$ is the length of the
longest proper chain of closed (or open) irreducible sets.
If $\cc$ is finite, then $d$ is finite, and at most the number of
objects minus one.
If $\cf$ as above is the set of
all nonidentity morphisms, then $\fleches{i}{\cc}$ is empty for
$i>d$; hence $\cc$ has homological dimension
at most $d$ (see \cite{gelfand}, meaning for any sheaves $F,G$ we have
${\rm Ext}^i(F,G)=0$ for $i>d$).

\subsection{Other Resolutions}
\label{sb:other_resolutions}

For an arbitrary finite, semitopological category, $\cc$, there are 
{\em greedy resolutions} of $F\in\myfield(\cc)$, constructed as follows.
For each $X\in\objects{\cc}$ we set
$$
G_X = \ker \bigoplus_{\phi\from Y\to X,\ Y\ne X}F\phi,
$$
in other words those elements of $F(X)$ that get sent to zero
by each $F\phi$
with $\phi$ having target $X$ and source not equal to $X$.
(If $X$ is an initial element, then $G_X=F(X)$.)
We choose an element in
$$
\iota_X\in\Hom(F(X),G_X)
$$
that restricts to the identity on $G_X$; we therefore get for each $X$ an
element of $\Hom(F,k_{X*}G_X)$.
This gives rise to a map
$$
\iota \from F\to I = \sum_{X\in\objects{\cc}} k_{X*}G(X)
$$
that we claim is initial in the category of inclusions of $F$ into an
injective; indeed,
let $\iota'\from F\to I'$ where $I'$ is a sum $k_{X*}V_X$;
note that we have a canonical isomorphism
$$
V_X\isom\ker \bigoplus_{\phi\from Y\to X,\ Y\ne X}I'\phi.
$$
Then $\iota'$ gives an injection $G_X\to V_X$, giving a morphism
$k_{X*}G_X\to k_{X*}V_X$, and a morphism $\nu\from I\to I'$.
We easily show that $\nu\circ\iota=\iota'$ and $\nu$ is uniquely determined
by this constraint, by structural induction on $\cc$ (i.e., we show
this at initial objects, and then remove the initial objects from
$\cc$, passing to the ``kernel'' of $F$ by $F\phi$ with $\phi$ having source
in an initial object (i.e., $i^!F$ with $i$ the closed inclusion),
and use induction).

By a {\em greedy} injective resolution of $F$ we mean any inductive resolution
$$
F\to I^0\to I^1\to\cdots
$$
with $I^j$ obtained greedily from the cokernel of the map to $I^{j-1}$.
Greedy resolutions are often much more efficient in practice (for example,
computing the cohomology of the examples in Section~3)
than the standard resolution.  

For reasons we do not understand, it is
often (but not always) the case that greedy resolutions of $\myfield$
and other sheaves (especially in ``geometric'' examples) have the
property that a summand $k_{X*}V_X$ for a fixed $X$ appears in only one
of the $I^j$ (in some sort of ``rank'' order).

{\em Greedy} projective resolutions can be defined similarly.

Next we describe a special resolution for finite, semitopological categories
isomorphic to a free category.
If $\cc$ is a free category, formed from the directed graph, $G$,
then the primes of $\cc$ are just the morphism corresponding to
$G$ edges.  We claim that 
any $F\in\myfield({\cc})$ has a projective resolution
$$
0\to P_1\to P_0\to F\to 0,
$$
where
$$
P_1 = 
\bigoplus_{\phi\in E(G)} k_{s(\phi)!} F\bigl(t(\phi)\bigr),\qquad
P_0 = \bigoplus_{X\in V(G)} k_{X!} F(X),
$$
where $V(G),E(G)$ are, respectively, the vertices and edges of $G$.
Indeed, it suffices to find for each $Y\in\objects{\cc}$
a chain homotopy, $K$, with 
$dK+Kd=\id$ on the above sequence localized at $Y$.  Note that
$$
P_0(Y)\isom \bigoplus_{X\in V(G)} F(X)^{\Hom(Y,X)},
$$  
so that a $P_0(Y)$ element, $w$, is equivalent to giving for each
$\nu\in\Hom(Y,X)$ (for any $X$) a ``component,'' $w^\nu\in F(X)$.
So let $K_{-1}\from F(Y)\to P_0(Y)$ be defined by mapping $v\in F(Y)$ to
$v$ in the component $F(Y)^{\id_Y}$ and zero elsewhere.  Clearly
$dK+Kd=\id$ on $F(Y)$.  Next note that
$$
P_1(Y)\isom \bigoplus_{\phi\in E(G)} 
F\bigl(t(\phi)\bigr)^{\Hom(Y,s(\phi))},
$$
so a $P_1(Y)$ element, $v$, is equivalent to specifying for each
prime $\phi$ (i.e., edge of $G$) and each $\mu\in\Hom(Y,s(\phi))$ a
``component,'' $v^{\mu,\phi}\in F(t\phi)$.
Set $K_0\from P_0(Y)\to P_1(Y)$ via
$$
(Kw)^{\mu,\phi}=\sum_{\alpha\in\morphisms{\cc}} 
(F\alpha)w^{\phi\circ\mu\circ\alpha}.
$$
Using unique factorization we easily verify that $dK+Kd=\id$ on $P_0(Y)$
and $P_1(Y)$.  For example, $Kd\from P_1(Y)\to P_1(Y)$ is given by
$$
(Kdw)^{\mu,\phi} = \sum_{\mu'\phi'\alpha=\mu\phi}(F\alpha)w^{\mu',\phi'}
- \sum_{\mu'\phi'\alpha=\mu}(F\alpha\phi)w^{\mu',\phi'},
$$
where $\alpha\in\morphisms{\cc}$ and $\phi,\phi'$ as before
(and notation $w^{\mu',\phi'}$ as before); the equation
$\mu'\phi'\alpha=\mu\phi$ can be solved by either $\alpha=\id$ or
$\alpha=\beta\phi$ for some $\beta$; the $\alpha=\id$ solution gives
$(Kdw)^{\mu,\phi}$ the summand $w^{\mu,\phi}$, where each solution where
$\alpha=\beta\phi$ gives rise to a solution $\mu'\phi'\beta=\mu$, causing
a cancellation.

\subsection{Left to Right, Right to Left, and Serre Functors}
\label{sb:serre}

Here we summarize the discussion of Appendix~\ref{ap:duality}, for
the special case used in this paper.

Let $\cd_1$ be a category whose $\Hom$ sets can be given the structure
of a finite dimensional
$\myfield$-vector space (see Appendix~\ref{ap:duality} for what this
entails; in this paper $\cd_1$ will be of the form
$\cd^{\rm b}(\myfield(\cc))$, the derived category
of bounded $\myfield(\cc)$ complexes, with $\cc$ finite and 
semitopological).  For $B\in\objects{\cd_1}$, we denote by
$(\LR)B$ or $B^{\LR}$, called {\em $B$ left-to-right}, the functor:
$$
A \mapsto \Hom(B,A)^*.
$$
If this functor is representable, we denote (by minor abuse of notation)
by $(\LR)B$ or $B^{\LR}$ any object representing.  If $B^{\LR}$ is
representable for any $B$, then Yoneda's lemma implies that
$(\LR)$ (called left-to-right)
extends to a (covariant) functor on $\cd_1$
(see Appendix~\ref{ap:duality}).
The left-to-right functor, if exists, has also been called the
{\em Serre functor} (see \cite{bondal_kapranov,bondal_orlov,larsen}, for 
example) in the context of the derived category.
(We prefer left-to-right and later $!\to *$ as names, since they are
more suggestive to our interests.)

The right-to-left functor is defined analogously, and is a pseudoinverse
of the left-to-right functor (when they are representable).
In this paper we shall give a simple construction of the Serre or
left-to-right functor for $\cd^{\rm b}(\myfield(\cc))$ as above.
We also remark that if $F,G$ are adjoints in $\cd_1,\cd_2$, i.e.,
$$
\Hom_{\cd_2}(FA,B)\isom \Hom_{\cd_1}(A,GB)
$$
(an isomorphism natural in $A,B$), then it is easy to see that
$G$ has a left adjoint
\begin{equation}\label{eq:lr_adjoint}
(\LR)_{\cd_2} F (\RL)_{\cd_1},
\end{equation}
provided the appropriate left-to-right and right-to-left functors
exist (see Appendix~A or \cite{larsen}, Remark~1.13).
So when left-to-right and right-to-left functors exist for both
categories, and adjoint pair $F,G$ can be extended indefinitely on the
left and right to a sequence of adjoints.

\subsection{Topological Spaces with a Group Action}

We claim that the topos of a
Grothendieck topology on a finite, semitopological category, $\cc$,
can be described as the category of $G$-invariant sheaves of sets
on a finite topological space, $X$, with a $G$ action, for some
(finite) group, $G$.  Indeed, we may assume $\cc$ has
its {\em grossi\`ere} topology.  The graph underlying $\cc$ has a
covering space (see \cite{geometric_aspects})
in which all multiple edges are separated in the cover;
by Galois theory of graphs
(see \cite{geometric_aspects}, for example) the covering graph has a
Galois covering\footnote{The proof of this theorem in \cite{geometric_aspects}
is incorrect; the following argument provides a correct proof.
If $H_1\to H_2$ is a covering map of graphs
of degree $n$, then the $n$-fold fiber product $H_1\times_{H_2}H_1\times_{H_2}
\cdots$ has its largest connected component of degree $n!$ over $H_2$,
and this component is a Galois cover.  See \cite{gross}.}.
This Galois cover inherits a composition law from $\cc$, and therefore
comes with the structure of a category, $\cc'$.  If $G$ is the Galois
group of the graph of $\cc'$ over that of $\cc$, then a sheaf of sets
on $\cc$ is the same thing as a $G$-invariant one on $\cc'$.  But
the underlying graph of $\cc'$ covers a graph with at most one edge
between any two vertices, so $\cc'$ is of topological type and yields
a topological space.

So, in a sense, we can always replace a finite, semitopological category
by a topological one with a finite group action.

\section{Examples of Categories}

A topological model involves a category, and we wish to give ways of
finding interesting examples of categories.

First we describe how interesting categories arise from geometry.
Say that a finite open covering $\{U_i\}$ of a topological space
or manifold, $M$,
is a {\em good cover} (respectively, {\em pretty good cover})
if (1) all $U_i$ have the cohomology of a
point (i.e., the same Betti numbers), and (2) each intersection $U_i\cap U_j$
equals some $U_k$ (respectively, is the union of some of the $U_k$'s).
(Our definition of good cover
is related to the nerve of a good cover in the sense
of \cite{bott}; our pretty good covers can be used to form hypercovers
as in [SGA4.V.7.3--4].)
It is known that good and pretty good covers
can be used to compute the Betti numbers of a space
(see [SGA4.V.7.3--4], \cite{bott,segal,dugger}, etc.).
In the following subsection we will study generalized coverings,
via {\espaces}, giving
categories that are not of topological type.

Second, in the subsection thereafter, we give specific
categories arising from manifolds
and some just arising from more combinatorial considerations.

Last we remark that there are general constructions to create new categories
out of old ones, such as the fiber product.

\subsection{Espaces \'Etal\'es}

Here we gather some facts on {\em espaces \'etal\'es}.

A morphisms $j\from X\to Y$ of topological spaces is an
{\em espace \'etal\'e (over $Y$)} 
iff it is a {\em local homeomorphism}, meaning
that for every $p\in X$ there is an open $U$ containing $p$ such that
$j(U)$ is open and $j|_{U}\from U\to j(U)$ is an isomorphism.  If
$j_i\from X_i\to Y$ for $i=1,2$ are two {\espaces}, then a morphism from
$j_1$ to $j_2$ is a morphism $\phi\from X_1\to X_2$ such that $j_2\phi=j_1$;
such a $\phi$ is necessarily an {\espace}.
Also the fiber product of $j_1$ and $j_2$ exists and is an {\espace}
$X_1\times_Y X_2\to Y$.

(The category of {\espaces} over $Y$ is equivalent to the category
of sheaves of sets over $Y$.)

If $U\subset X$ is an open subset of a topological space, $X$, we write
$j_U$ for the inclusion $U\to X$; $j_U$ is 
an {\espace}.

If $j\from X\to Y$ is an {\espace}, and $U\subset X$ is open, then
$j(U)$ is open.  The pullback, $j^*$, acting on sheaves of $\myfield$-vector
spaces over $Y$ to those on $X$, has an exact left
adjoint, $j_!$, given as the sheaf associated to the presheaf
\begin{equation}\label{eq:j_shriek}
(j_!^{\rm pre}F)(U)= \bigoplus_{\phi\from j_U\to j} F(\phi(U))
\end{equation}
(see [SGA4.IV.11.3.1]\footnote{Proving that the sheaf associated to
the presheaf in equation~(\ref{eq:j_shriek}) really is the left adjoint
to $j^*$ makes a nice exercise.  Indeed,
an element of $\Hom(j_!F,G)$ gives a
Hom from the direct sum of $F(\phi(U))$'s to $G(U)$, or equivalently a
product of the individual Homs; if $V\subset X$ is open with
$j|_V\from V\to j(V)$ an isomorphism, then $j^*G(V)=V\times_Y G=
j(V)\times_Y G =G(j(V))$ and
we have a $\phi\from j_{j(V)}
\to j$ with $\phi(j(V))=V$, giving an element of $\Hom(F(V),G(j(V)))=
\Hom(F(V),j^*G(V))$.  We need to check that these $\Hom(F(V),j^*G(V))$
agree on overlaps, and that the resulting map from Hom sets is
a functorial bijection.
}).  
It easily follows that the stalk $(j_!F)_p$ (or, equivalently, 
$(j_!^{\rm pre}F)_p$) is isomorphic to
$$
\bigoplus_{q\ {\rm s.t.}\ j(q)=p} F_q .
$$


Since $j_!$ has an exact right adjoint (namely $j^*$), $j_!$ takes
projectives to projectives.  It follows that
$$
{\rm Ext}^i(j_!F,G) = {\rm Ext}^i(F,j^*G).
$$

\subsection{Finite Categories Arising from Topological Spaces}

In this subsection we wish to describe how finite categories can arise
from topological spaces in a natural way such that the cohomology of
the category agrees with that of the space.
We shall give a theory that, among other things,
gives categories that are not of topological type.  Let us start with an
example.

Consider the open real intervals $U_1=(-0.1,0.1)$ and $U_2=(-0.1,1.1)$.
If $S^1=\reals/\integers$, then the natural map $\reals\to S^1$ induces
natural
maps $\iota_i\from U_i\to S^1$.  There are two maps from $\iota_1$
to $\iota_2$ (namely addition by either $0$ or $1$).  A sheaf on $S^1$
pulls back to one on $U_2$, and a sheaf on $U_2$ arises as a pullback
precisely when its two pullbacks to $U_1$ agree.  We claim that this
fact and the fact that $U_1$ and $U_2$ are both contractible implies
that the cohomology of $S^1$ agrees with that of $\cu$, where
$\cu$ is the category whose objects are $\{\iota_1,\iota_2\}$ and with
two morphisms from $\iota_1$ to $\iota_2$ (and those are the only
nonidentity morphisms).  We
shall give a general principle to this effect.

Let
$$
\cdots\threerightarrows M_1\tworightarrows M_0\rightarrow M_{-1}=M
$$
be a simplicial topological space, with all arrows being {\espaces}.
Let $j_i\from M_i\to M$ be the composite arrow.
We claim that the two conditions:
\begin{equation}\label{eq:cond_exact}
\mbox{
$\cdots\rightarrow j_{1!}\myfield \rightarrow j_{0!}\myfield\rightarrow
\myfield$ \ \ is exact,}
\end{equation}
and
\begin{equation}\label{eq:acyclic}
\mbox{ $H^j(M_i,\myfield)=0$ for all $j\ge 1$ and all $i\ge 0$,}
\end{equation}
imply that $H^i(M,\myfield)$ is the $i$-th cohomology group of
\begin{equation}\label{eq:degenerate_sequence}
0\rightarrow H^0(M_0,\myfield)\rightarrow H^0(M_1,\myfield)\rightarrow\cdots
\end{equation}
This follows by using Condition~(\ref{eq:cond_exact}) in the first
variable of
${\rm Ext}^*(\myfield,\myfield)$ (which is $H^*(M,\myfield)$) to
obtain a degenerate spectral sequence that degenerates
to equation~(\ref{eq:degenerate_sequence}).

Next let us specialize to the case where there is a category, $\cu$,
and a topological space $M=M_{-1}$ with the following data.  To each
$X\in\objects{\cu}$ there corresponds an {\espace},
$\iota_X\from M_X\to M$.
By an {\em correspondence \'etal\'e} from $\iota_X$ to $\iota_Y$ we mean 
{\espaces} $\eta_1\from N\to M_X$ and
$\eta_2\from N\to M_Y$ such that $\iota_X\eta_1=\iota_Y\eta_2$;
we shall abbreviate this as $\eta\from N\to M_X\times_M M_Y$ and
speak of following $\eta$ with projections ${\rm pr}_1,{\rm pr}_2$,
respectively, to obtain $\eta_1,\eta_2$, respectively; since 
$\iota_X,\iota_Y$ are {\espaces}, $M_X\times_M M_Y$ actually exists.
Similarly we define an {\em $r$-correspondence \'etal\'e} on a tuple
$\iota_{X_1},\ldots,\iota_{X_r}$ via a map
$$
\eta\from N\to M_{X_1} \times_M \cdots \times_M M_{X_r}.
$$
\begin{definition} Let $\cu$ be a category and $M$ a topological
space.  By a $\cu$-covering of $M$, $\iota$, we mean the data consisting of
(1) for each $X\in\objects{\cu}$ an {\espace} $\iota_X\from M_X\to M$,
and (2) for each $\phi\in\morphisms{\cu}$ a correspondence \'etal\'e
$\iota_\phi\from M_\phi \to M_{s\phi}\times_M M_{t\phi}$.
\end{definition}

Consider a $\cu$-covering of $M$, $\iota$.
For each composable sequence $\vec\phi=(\phi_1,\ldots,\phi_r)$ in $\cu$,
we can compose correspondences as usual to get an $r$-correspondence
$$
\iota_{\vec\phi}\from M_{\vec\phi} \to 
M_{s\phi_1}\times_M\cdots \times_M M_{s\phi_r}\times_M M_{t\phi_r}.
$$
where
$$
M_{\vec\phi}=
M_{\phi_1}\times_{M_{t\phi_1}} \cdots
\times_{M_{t\phi_{r-1}}} M_{\phi_r}.
$$
We get a simplicial space \'etal\'e by setting
$$
M_j = \coprod_{\phi\in\fleches{j}{\cu}} M_\phi.
$$
\begin{definition} We say that a $\cu$-covering of $M$, $\iota$, is
{\em cohomologically faithful} if for each $j\ge 0$ and
$\vec\phi\in\fleches{j}{\cu}$ we have that $M_{\vec\phi}$ has the
cohomology of a point (i.e., one dimensional in degree $0$ and vanishing
in higher degrees).
\end{definition}
In this case the cohomology of
equation~(\ref{eq:degenerate_sequence}) is the cohomology
of $\cu$ (using the standard projective resolution of $\myfield$ in $\cu$).
We have seen the following theorem.

\begin{theorem} Consider a cohomologically faithful $\cu$-covering of
$M$, $\iota$, for which condition~(\ref{eq:cond_exact}) holds.  Then
the cohomology of $M$ is that of $\cu$.
\end{theorem}
To check condition~(\ref{eq:cond_exact}) it suffices to check the stalks.
Let $\cm_p$ be the category whose objects are $j_0^{-1}(p)$ (recall
$j_i$ is the {\espace} $M_i\to M$) and morphisms are $j_1^{-1}(p)$
and source and target maps come from the correspondence.  
We have $j_i^{-1}(p)$ is the set of composable morphisms of length $i$
in $\cm_p$.
The following
theorem follows immediately.
\begin{theorem} Consider a $\cu$-covering of $M$, $\iota$.  
With notation as above, let each $\cm_p$ be finite.
Then condition~(\ref{eq:cond_exact}) holds iff each $\cm_p$ has the
cohomology of a point.  (Note that the empty category does not have
the cohomology of a point.)
\end{theorem}

We give some common practical situations.
If each for each $\phi\in\morphisms{\cu}$,
$\iota_\phi$ is a morphism $\iota_{s\phi}\to\iota_{t\phi}$,
then $M_\phi=M_{s\phi}$ and we are in the following situation.
\begin{theorem} Consider a $\cu$-covering of $M$, $\iota$, for which
the composition
$$
\begin{CD}
M_\phi@>{\iota_\phi}>>  M_{s\phi}\times_M M_{t\phi} @>{{\rm pr}_1}>>
M_{s\phi}
\end{CD}
$$
is an isomorphism.
Then for $\vec\phi=(\phi_1,\ldots,
\phi_k)$ we have $M_{\vec\phi}\isom M_{s\phi_1}$.  In particular, in
such a situation we have that $\iota$ is cohomologically faithful provided
that for each $X\in\objects{\cu}$, $M_X$ has the cohomology of a point.
\end{theorem}

\begin{definition}
By a {\em $\cu$-quasihypercover} of $M$, $\iota$,
we mean a $\cu$-cover such that
(1) for each $\phi\in\morphisms{\cu}$,
$\iota_\phi$ is a morphism $\iota_{s\phi}\to\iota_{t\phi}$,
(2) each $p\in M$ is in the image of some $\iota_X\from M_X\to M$
for an $X\in\objects{\cu}$, 
(3) all the $\iota_\phi$ are
distinct, (4) for each
$X,Y\in\objects{\cu}$, we have that $M_X\times_M M_Y$ is covered
by the correspondences, meaning that it is the union of images
$\iota_Z\to \iota_X\times\iota_Y$ of a product of two correspondences.
\end{definition}
\begin{theorem} Let $\cu$ be a semitoplogical category (not necessarily
finite).  Let a topological space, $M$, 
admit a $\cu$-quasihypercover, $\iota$, 
such that
each $M_X$ has the cohomology of a point.  Then the cohomology of 
$M$ and $\cu$
agree.
\end{theorem}
\proof 
It suffices to show condition~(\ref{eq:cond_exact}), i.e., that for any
fixed $p$ the sequence
\begin{equation}\label{eq:local_exact}
\begin{CD}
\cdots @>{d_1}>> \bigoplus_{\phi\in\fleches{1}{\cm_p}}\myfield 
@>{d_0}>> \bigoplus_{q\in\fleches{0}{\cm_p}}\myfield@>{d_{-1}}>>
\myfield @>>> 0
\end{CD}
\end{equation}
is exact.  Note that $\cm_p$ is nonempty, by condition~(2) of the
definition of quasihypercover, and thus $d_{-1}$ is surjective.

First we claim that each $\cm_p$ is of topological type.  Indeed,
assume not.  Then there exist $\mu_1,\mu_2$, distinct
morphisms $\iota_X\to\iota_Y$ for objects
$X,Y\in\objects{\cu}$, such that for $q\in M_X$ with $\iota_X(q)=p$
we have $\mu_1(q)=\mu_2(q)$.  The set where $\mu_1=\mu_2$ is closed
(in any topological setting) and open (since they are {\espaces}),
contains a point (namely, $q$), but is not all of $M_X$; therefore $M_X$
has at least two connected components and does not have the cohomology
of a point.

For each
$q_1,q_2\in\objects{\cm_p}$, 
condition~(4) on quasihypercovers shows that there is a
$q_3\in\objects{\cm_p}$ with arrows to both $q_1$ and $q_2$.
In particular, if $\cm_p$ is finite then it has an initial element.

Assume that $\cm_p$ is finite.
Any category of topological type with an
initial element has the cohomology of a point (using the fact that
$\myfield$ is injective, equalling $j_*\myfield$ where $j$ is the
inclusion of the initial element of the category into the category).
So we are done.

If $\cm_p$ is not finite, any element, $\eta$, of a direct sum in
equation~(\ref{eq:local_exact}) vanishes on all but finitely many
components, and is therefore supported on a finite,
full subcategory $\cm'_p$ of
$\cm_p$.  If $\cm'_p$ has $k$ objects, $X_1,\ldots,X_k$, set $Y_1=X_1$ and
let $Y_i$ be inductively defined for $i\ge 2$ as an element with an arrow
to $X_i$ and $Y_{i-1}$.  Let $\cm''_p$ be the full subcategory of $\cm_p$
on the objects $X_1,Y_1,\ldots,X_k,Y_k$.
The $\cm''_p$ is finite, has an initial element (namely $Y_k$), and
supports $\eta$; as mentioned before, this means $\cm''_p$ has the 
cohomology of a point.  Hence if $\eta\in\ker d_i$, then it is in the
image of $d_{i+1}$ (first restricting to $\cm''_p$ and then extending
by zero to $\cm_p$). 
\proofbox


We remark that this theorem does not cover all interesting cases.
The following class of examples is joint with Denis Sjerve.
Consider again $M=M_{-1}=\reals/\integers$, $M_0=U_1\amalg U_2$ with
$U_2=(-.1,3.1)$ and $U_1=(-.1,2.1)$ mapped naturally to $M$.
We set $M_1$ to be two copies of $U_1$, representing the two maps
addition by $0$ and by $1$.
$U_2\times_M U_2$ consists of a number of ``strips,'' i.e., pairs $(a,b)$
in $U_2\times U_2$ with $a-b\in\integers$, but only the longer strips
are ``covered'' by $U_1$ (that covers $a-b=\pm 1$) and $U_2$ (that covers
$a=b$).  So this is not a quasihypercover.  
On the other hand, each $\cm_p$
has the cohomology of a point, as do the $U_i$'s, and so the cohomology
of the resulting category is that of $M$.

(More interesting examples can be obtained with $\reals/\integers$ 
covered by more and different size intervals.)




\subsection{Examples}

Here we give some examples of categories.

Consider the boundary of a simplex on $n$-vertices, and
extend each of its $n$ faces (of dimension $n-1$) slightly to an open
set.  The resulting category is $(\cat{1})^n$ minus its initial object and
its terminal object; this category and morphism therefore models 
the sphere $S^{n-1}$.  (This corresponds to a good cover; the quasihypercover
consists of the set of intersections of any of the extended faces.)

Cover the sphere $S^n$ with an open upper and lower ``extended''
hemispheres (each
extending past the ``equator''), covering
the hemisphere intersection $\isom \reals\times S^{n-1}$ by covering
$S^{n-1}$ with extended hemispheres, etc., we see that $S^n$ admits a
quasihypercover associated
to a category whose objects are $\{U_0,L_0,\ldots,U_n,L_n\}$
with inclusions as the object index is increased. 
(This is comes from a
pretty good cover, where the intersection of hemispheres
can have nontrivial cohomology.)

As mentioned before, a line segment that meets itself at its ends 
gives rise to a
category with objects $\{M_0,M_1\}$, with two morphisms from $M_0$ to $M_1$,
corresponding to the two inclusions of the self-intersection segment in
the circle into the line segment.  Another way to achieve this is to
act on the category in the previous paragraph by the cyclic group of order
two, corresponding
to the antipodal map on the sphere.  We conclude that real projective
$n$-space (the quotient of $S^n$ by the antipodal map) is modelled
by the category
with objects $\{M_0,\ldots,M_n\}$, with two morphisms between objects
of increasing index which can be labelled $\{+,-\}$
such that composition given by
multiplying signs.

We finish with some general (less geometric) remarks on finite categories.
$\cat{1}$ can be viewed as the tautological open/closed pair, in that
to given an open (or closed) set in $\cc$ is the same as to give a 
morphism $\cc\to\cat{1}$.  Similarly $(\cat{1})^n$ is the tautological
ordered $n$-tuple of open/closed pairs.
New categories can be obtained from old ones by limits; we shall be
especially interested in the fiber product.

An``$(m+1)$-partite'' category is a category of topological type,
$\cc$,
with objects consisting of $m+1$ sets $S_0,\ldots,S_m$, with
$|S_i|=n_i$, and with respectively one or zero morphism
from an object in $S_i$ to one in $S_j$ according to whether or not
$i\le j$.  The greedy resolution easily shows that
$$
h^i(\cc,\myfield) = \left\{ \begin{array}{ll} 1 & \mbox{if $i=0$,} \\
 (n_0-1)\ldots(n_m-1) & \mbox{if $i=m$,} \\
 0 & \mbox{otherwise.} \end{array}\right.
$$
The standard resolution shows that the $m$-th
Betti number of $\cc$ can be bounded
by the number of composable $m$-tuples not containing an identity.  The
above computation for $\cc$ shows that this bound can be ``close'' to true
(at least for $m$ fixed and all $n_i$ ``large''),
since the number of such tuples is $n_0\ldots n_m$.

\section{Virtual Zero Extensions}
\label{se:zero_extensions}

Consider equation~(\ref{eq:and}).
Assume that $U_{f\wedge g}=U_f\cap U_g$ for all $f,g$.
Let $Z_g$ be the closed complement to $U_g$, for each $g$.
The exact sequence
$$
0\to G_{U_g} \to G \to G_{Z_g}\to 0,
$$
shows that ${\rm cc}(F,G_{Z_g})$ is within ${\rm cc}(F,G)$ of
${\rm cc}(F,G_{U_g})$.  
\begin{definition} Let $G\in\myfield(\cc)$.  Let $U,Z$ respectively be
an open and a closed set in $\cc$.  A {\em virtual $G_{U,Z}$} (or
{\em virtual zero extension for $G,U,Z$}) is a sheaf
$H$ and arrows $G_U\to H\to G_Z$ such that
$$
0\to G_{U} \to G_{U\cap Z}\oplus H \to G_{Z}\to 0;
$$
is exact, where the map to $G_{U\cap Z}$ is the identity on $U\cap Z$
and the map from $G_{U\cap Z}$ is minus the identity on $U\cap Z$.
\end{definition}
Virtual $G_{U,Z}$'s form a category, with a morphism from $H_1$ to $H_2$
(with arrows from $G_U$ and to $G_Z$) being $\mu\from H_1\to H_2$ such that
$$
\begin{CD}
G_U @>>>  H_1 @>>> G_{Z} \\
 @VV{\id}V   @VV{\mu}V  @V{\id}VV \\
 G_U @>>> H_2 @>>> G_{Z}
\end{CD}
$$
commutes everywhere.
Notice that for $G$ as above and any full subcategory, $A$, of $\cc$,
we can define $G_A$, the ``literal extension of $G$ on $A$ by $0$,''
to be the sheaf that agrees with $G$ on $A$ and is extended by $0$ outside
of $G$ if this sheaf exists; the issue in existence 
is that if $\phi\in\morphisms{\cc}$ factors through an element
outside of $A$, then $G_A\phi$ is forced to $0$, which creates a conflict
if $G\phi$ is not zero.  
If $G_A$ exists and $A=U\cup Z$ with $U$ open, $Z$ closed, then
$G_A$ is a virtual $G_{U,Z}$.

The following theorem is easy to check.
\begin{theorem} Let $\cc$ be a finite category of topological type, and 
$A\subset\objects{\cc}$.  The following are equivalent:
\begin{enumerate}
\item $A$ is an {\em open/closed intersection}, i.e., the intersection of an
open set with a closed set,
\item $A$ is the intersection of ${\rm open}(A)$ with ${\rm closed}(A)$,
where ${\rm open}(A)$ is the smallest open set containing $A$ and
similarly for ${\rm closed}(A)$,
\item $A$ is {\em cavity-free}, i.e., if $\phi_1,\phi_2$ are composable
morphisms with $s\phi_1,t\phi_2\in A$, then $t\phi_1=s\phi_2\in A$,
\item for all $F\in\myfield(\cc)$ we have 
that the literal zero extension $F_A$ exists.
\end{enumerate}
\end{theorem}
Notice that for $\cc=\cat{2}$ and $A={0,2}$, the conditions of the theorem
do not hold, and yet a virtual $F_{\{0\},\{2\}}$ always exists.


If $U\cap Z=\emptyset$ (with $U$ open, $Z$ closed), then the category of
virtual $G_{U,Z}$'s is the same as the Yoneda ${\rm Ext}(G_Z,G_U)$
category.

Virtual extensions always exist in free categories
(that don't have the type of conflict described earlier, since each
morphism has a unique factorization).

Let us make some structural observations.
\begin{definition} A virtual $G_{U,Z}$, $H$, is {\em standard} if
(1) $H(X)$ is $G(X)$ or $0$ according to whether or not
$X\in U\cup Z$, (2) $H\phi=0$ if $s\phi$ or $t\phi$ lies outside
$U\cup Z$, (3) $H\phi=G\phi$ if $s\phi$ and $t\phi$ both lie in
$U$ or both in $Z$.
\end{definition}
For a standard virtual $G_{U,Z}$, $H$, the only morphisms $H\phi$ that
are not determined are those with $s\phi\in U\setminus Z$ and 
$t\phi\in Z\setminus U$.

\begin{theorem} In the category of virtual $G_{U,Z}$'s, each isomorphism
class contains exactly one standard virtual $G_{U,Z}$.  The virtual
$G_{U,Z}$'s form a $\myfield$-vector space, via (a modification of)
the Baer sum; when we restrict this sum to standard virtual $G_{U,Z}$'s,
the vector space structure is given by mapping $H$ to the sum indexed over 
$\phi\in\Hom(X,Y)$ with $Y\in Z\setminus U$ and $X\in U\setminus Z$ 
of $H\phi\in\Hom(G(Y),G(X))$ (and we may restrict to $\phi$ prime if we
like).
\end{theorem}
\proof 
If for each $X\in\objects{\cc}$ we have isomorphisms
$\iota_X\from H(X)\to t\iota_X$, then we define the {\em conjugate
of $H$ by $\iota$} to be the sheaf, $H'$, such that
$H'(X)=t\iota_X$ and $H'\phi=\iota_{s\phi}(H\phi)\iota_{t\phi}^{-1}$.
Note that $H'$ is in the same isomorphism class as $H$, with
$\iota$ giving rise to an isomorphism $H\to H'$.

The exact sequence
$$
0\to G_U\to G_{U\cap Z}\oplus H\to G_Z\to 0
$$
shows that for $X\in U\setminus Z$ we may choose an isomorphism
$\iota_X \from H(X)\to G(X)$.  
Similarly for $X\in Z\setminus U$.  For $X\in U\cap Z$, we have
$$
0\to G(X)\xrightarrow{\id\oplus\alpha} G(X)\oplus H(X)
\xrightarrow{(-\id)\oplus\beta} G(X)\to 0.
$$
We have $\beta\alpha=\id$; set $\iota_X=\alpha$.  Then we easily check that
$H$ conjugated
by $\iota$ gives rise to a standard virtual $G_{U,Z}$.

If $G_U\xrightarrow{\alpha_i}H_i\xrightarrow{\beta_i}G_Z$ 
for $i=1,2$ are two virtual $G_{U,Z}$'s, for each $X\in\objects{\cc}$
we consider pairs $(h_1,h_2)$, with $h_i\in H_i(X)$ such that
$\beta_1(h_1)-\beta_2(h_2)$ agree on $Z\setminus{U}$; let $H_3(X)$ be
the set of such pairs modulo the image of $\alpha_1\oplus\alpha_2$.
(This construction comes from the Baer sum, $G_{U\cap Z}\oplus H_i$
being an extension of $G_Z$ by $G_U$.)
We easily check $H_3$ is a virtual $G_{U,Z}$.  Furthermore, if $H_1,H_2$
are standard, then so is $H_3$, and for $\phi$ with $t\phi\in Z\setminus U$   
and $s\phi\in U\setminus Z$ we have 
$$
H_3\phi = H_1\phi+H_2\phi.
$$
Finally, the $H\phi$ as above are determined by $H\phi$ for $\phi$ prime,
since any factorization of such a $\phi$ contains exactly one morphism
with source in $U\setminus Z$ and target in $Z\setminus U$.
\proofbox

Assume for each $f,g$ there is a virtual $G_{U_f,Z_g}$.
Then the resulting short exact sequence gives
$$
{\rm cc}(F,G_{U_f\cap Z_g})\le {\rm cc}(G_{U_f})+{\rm cc}(G_{Z_g}).
$$
The exact sequence
$$
0\to G_{U_f\cap U_g} \to G_{U_f} \to  G_{U_f\cap Z_g} \to 0,
$$
gives
$$
{\rm cc}(F,G_{U_f\cap U_g})\le {\rm cc}(G_{U_f})+{\rm cc}(G_{U_f\cap Z_g}).
$$
We conclude
$$
{\rm cc}(f\wedge g) \le 2\ {\rm cc}(f) + {\rm cc}(g) + {\rm cc}(F,G).
$$

Virtual zero extensions exist in the following two extreme cases:
(1) each $U_f$ is also closed, and
(2) $\cc$ is a free category.  The problem with the first case
is all the $U_f,Z_f$'s are disconnected, and the bounds are trivial.
The problem with the second is that $\widehat{\cal C}$ is homologically
one dimensional (see Subsection~\ref{sb:other_resolutions}), and we
think it less likely that sheaf models based on such $\cc$ will give
interesting bounds.  For example,
all cohomology boils down to $H^0$ and the Euler characteristic; $H^0$ is
usually simple to determine, and the Euler characteristic has
the simple formula:
$$
\chi(G) = \sum_{P\in\objects{\cal C}} \dim\bigl(G(P)\bigr) - 
\sum_{\phi\;\;{\rm prime}} \dim\Bigl(G\bigl({\rm source}(\phi)\bigr)\Bigr).
$$

It is possible to give some simple variants on these ideas.  For example,
one could find a ``near zero extension,'' i.e., an $H$
such that the derivation of (the non-exact in the middle):
$$
0\to G_{U} \to G_{U\cap Z}\oplus H \to G_{Z}\to 0
$$
is not nonzero, but has, say, a middle term, $M$; then we simply add
${\rm cc}(F,M)$ appropriately into the bounds.

Let us note that virtual zero extensions are preserved under pulling back.
In other words, if $u\from\cc'\to\cc$ is an arbitrary functor, and we
have that $H$ is a virtual $G_{U,Z}$ (all, as before, over $\cc$), then
$u^*H$ is a virtual $(u^*G)_{u^{-1}(U),u^{-1}(Z)}$.
Furthermore, say that there is a virtual $G_{U,Z}$ {\em in $\cc'$} (with
$u\from\cc'\to\cc$ understood, or say {\em in $u$}) if
there is a virtual $(u^*G)_{u^{-1}(U),u^{-1}(Z)}$.  So each time we
pullback to a category, vitual zero extensions are never destroyed
and new ones may be created (although the notion of the zero extension
depends, of course, on how the sheaf and open and closed sets pullback).

This suggests a possible fiber product construction.
Say that for each $f,g$ we choose a $u=u_{f,g}\from\cc_{f,g}\to
\cc$ such that
if there is a virtual $G_{U_f,Z_g}$ in $\cc_{f,g}$.
The fiber product, $\cx$, of the $\cc_{f,g}$'s over $\cc$ has 
a virtual $G_{U_f,Z_g}$ in $\cx$ for any $f,g$.
Of course, such a construction, given that there are $2^{2^n}$
possible $f$'s and possible $g$'s, would yield a large category.

It is for this reason that we study the behavior of duality (which we
have in mind for negation) under fiber products (and therefore, more
generally, under arbitrary base change).

In the sections to follow we will see that it may be possible
to maintain a reasonable duality theory while performing fiber product
operations.

\section{Injectives and Projectives}
\label{se:injectives}
In this section we describe the structure of injective and projective
modules in $\vs{\cc}$ for a finite, semitopological category, $\cc$.
Then we describe a naturally arising map $*\to !$ and its quasi-inverse
$!\to *$ (defined in the derived category); 
intuitively, $*\to !$ is constructed by taking a complex
of sheaves, writing an injective resolution, writing each injective as
a sum of modules $(k_X)_*V$ (see below), and replacing the $*$ with a
$!$.
Then we define the trace of a map, either from an injective or projective
to itself, or from an injective, $I$, to $I^{*\to !}$, or from a projective,
$P$, to $P^{!\to *}$; this trace is constructed using the structure
of injectives and projectives.

Let $\cc$ be a finite
semitopological category, and let $\ci$ be the injectives
of $\vs{\cc}$.  We begin by describing $\ci$.

If $X\in\objects{\cc}$, we set $k_X\from \Delta_0\to \cc$ to be the inclusion 
functor mapping $0$ to $X$.  Let $\cv=\vs{\Delta_0}$ be the category of 
vector spaces.  For $V\in\cv$ we have $(k_X)_*V$ is injective.

\begin{theorem}\label{th:injsum}
Every element of $\ci$ is the direct sum of injectives
of the form $(k_X)_*V$.  More precisely, an $I\in\ci$ is the sum of
$(k_X)_*V_X$, where for each $X$, if $\phi_1,\ldots,\phi_r$ are the
morphisms with target $X$, then
\begin{equation}\label{eq:injsum}
V_X = \ker(I\phi_1\oplus\cdots\oplus I\phi_r).
\end{equation}
\end{theorem}
\proof We prove the theorem by induction on the number of objects in
$\cc$.
Let $X$ be an object such that $I(X)$ is nonzero and
$X$ is minimal with this property, i.e., if $\phi$ is a morphism with
target $X$ then $I$ is zero on the source of $\phi$.  Let $k=k_X$,
$V=I(X)$, and $G=k_*V$; let $G_X=k_!V$ be the sheaf that is zero outside
$X$ and with $G(X)=V$.  The inclusion of $G_X$ into $G$ and the map from
$G_X$ to $I$ gives rise to a map $\psi\from G\to I$.

We claim that for any $Y$ we have $\psi(Y)\from G(Y)\to I(Y)$ is an injection
(and therefore $\psi$ is an injection).
Indeed, let $\phi_1,\ldots,\phi_s$ be the morphisms from $X$ to $Y$.
We have a commuting diagram
$$
\begin{array}{ccc} G(Y) & \xrightarrow{G\phi_1\oplus\cdots\oplus G\phi_s} 
& G(X)^s \\
\downarrow & &\downarrow \\
I(Y) & \xrightarrow{I\phi_1\oplus\cdots\oplus I\phi_s} & I(X)^s \end{array}
$$
The top arrow is an isomorphism, as is the right arrow.  Hence the left
arrow is an injection, which was the claim.

Hence $\psi\from G\to I$ is an injection.  A standard argument shows
that the cokernel of an injection of injectives is injective
(see, e.g., \cite{gelfand}), and thus
the cokernel $I/G$ is also injective, and hence $I$ is a direct
sum of $G$ and $I/G$.  
Thus equation~(\ref{eq:injsum}) holds (for that particular $X$).
Let $\cc'$ be the full subcategory of $\cc$ with $X$ removed.
$I/G$ vanishes at $X$, and so $I/G$'s restriction to $\cc'$ is injective;
by induction we have that
$I/G$ restricted to $\cc'$ is a direct sum as above; the same is true
viewing $I/G$ on $\cc$ (extended by $0$ on $X$).
For any $Y$, let $\phi_1,\ldots,\phi_s$ be the morphisms from $X$ to $Y$.
Then $I\phi_1\oplus\cdots\oplus I\phi_s$ is an injection on $G(Y)$ and
vanishes on $(I/G)(Y)$.  This shows equation~(\ref{eq:injsum}) with
(arbitrary) $Y$ replacing $X$,
given that it holds for $I/G$ on $\cc'$. 
\proofbox


Let $\cc$ be a finite semitopological category and $\cv$ be the
category of all finite dimensional vector spaces, with notation as
before.  Let
$\nu\from \simexp{\cc}{\cv}\to \ci$ be the functor given as follows;
for $F\in\Hom\bigl(\objects{\cc},\objects{\cv}\bigr)$ set
$$
\nu(F) =
\bigoplus_{X\in\objects{\cc}} (k_X)_* F(X).
$$
Notice that
$$
\Hom\bigl( (k_X)_* F(X), (k_Y)_* G(Y) \bigr) \isom
\Hom\bigl( k_Y^*(k_X)_* F(X), G(Y) \bigr)
$$
$$
 \isom
\bigoplus_{\phi\from X\to Y} \Hom\bigl(F(X),G(Y)\bigr);
$$
A morphism, $H\from F\to G$, in $\simexp{\cc}{\cv}$ gives for each
$\phi\from X\to Y$ an element of $ \Hom\bigl(F(X),G(Y)\bigr)$, and
therefore an element of 
$$
\Hom\bigl( (k_X)_* F(X), (k_Y)_* G(Y) \bigr)
$$
for each $X$ and $Y$, and therefore a morphism in $\ci$.
We easily verify that this makes $\nu$ a functor.

\begin{theorem}
\label{th:fullyfaithful}
The functor $\nu$ defines an equivalence of
categories between $\simexp{\cc}{\cv}$ and $\ci$; in other words,
$\nu$ is fully faithful and essentially surjective.
\end{theorem}
\proof 
Theorem~\ref{th:injsum} shows that $\nu$ is essentially surjective.
Fix objects $F,G$ of $\simexp{\cc}{\cv}$.
A morphism from a direct sum to another direct sum decomposes into
morphisms from each direct summand to each in the other.
Thus there is a one-to-one correspondence between elements of
$\Hom(\nu F,\nu G)$ and the direct sum over $X,Y$ objects of $\cc$ of
$$
\Hom\bigl( (k_X)_* F(X), (k_Y)_* G(Y) \bigr) \isom
\bigoplus_{\phi\from X\to Y} \Hom\bigl(F(X),G(Y)\bigr).
$$
But the right-hand-side summed over all $X,Y$ just gives $\Hom(F,G)$;
this gives a one-to-one correspondence between $\Hom(F,G)$ and
$\Hom(\nu F,\nu G)$).
Hence $\nu$ is fully faithful.
\proofbox
Henceforth with denote by $\nu_{\to *}$
the functor $\nu$ in the above theorem.

Now we wish to describe a certain class of quasi-inverses to $\nu_{\to *}$
that we will use.  Given $I\in\objects{\ci}$ we define
$$
(\mu I)(X)=
\ker \left( 
\bigoplus_{t(\phi)=X}
I\phi \right)
$$
(as in equation~(\ref{eq:injsum})).  It is easy to see that for
each $F\in\objects{\simexp{\cc}{\cv}}$, for each $X\in\objects{\cc}$ we have
that $F(X)\isom\mu\nu_{\to *}F(X)$.  
In a sense $\mu$ comes close to being a quasi-inverse;
an $f\from I_1\to I_2$ clearly determines a map
$$
(\mu f)({\rm Id}(X))\from (\mu I_1)(X)\to (\mu I_2)(X).
$$
Thus we may speak of $\mu$ as a map on objects and ``diagonal parts
of morphisms.''
However, given $\phi\in\morphisms{\cc}$ with $\phi\from X\to Y$ and
$X\ne Y$, there seems to be no canonical choice for a morphism
$$
(\mu f)\phi\from (\mu I_1)(X)\to (\mu I_2)(Y).
$$
(For example, try to make a canonical choice in the case where
$\cc=\cat{1}$, $X=0$, $Y=1$, and $I_1,I_2$ are both isomorphic to $J$,
where $J(0)\isom\myfield$ and $J(1)\isom\myfield^2$ (therefore
$J(1)\to J(0)$ is surjective, since $J$ is injective); 
how can an $f\from I_1\to I_2$ determine a map from $(\mu I_1)(0)$ to 
$(\mu I_2)(1)$?)

\begin{definition} By the {\em ``to star'' functor} we mean the
functor $\nu_{\to *}$ above.  By the {\em ``from star'' functor},
$\nu_{*\to}$ we mean any quasi-inverse that agrees with
$\mu$ as a map on objects and on diagonal parts of morphisms.
\end{definition}
In other words, for each $I\in\ci$ we choose an isomorphism
$\iota\from \nu_{\to *}\mu I\to I$ such that $\mu$ maps $\iota$ to
the identity map on the diagonal parts of $\iota$; such an $\iota$
exists by the fully faithfulness of $\nu_{\to *}$.  Then the choice of
such an $\iota$ for each $I$ determines, as usual, a quasi-inverse
$\nu_{*\to}$.

Note that the Axiom of Choice implicit in the last paragraph is not
really necessary if we are interested in applying it
(in practice) to only finitely many objects (see 
Section~\ref{sb:axiom of choice}).

Let us mention that our restriction to a special type of quasi-inverse,
$\nu_{*\to}$, will make the definition of a certain trace independent
of the choice of quasi-inverse; see below.  
(However, it is not clear to us that
this independence is absolutely necessary.)

By duality, and in particular by replacing $*$ with $!$,
we get a similar functor 
(called ``to shriek'') $\nu_{\to !}\from \simexp{\cc}{\cv}\to\cp$
where $\cp$ is the category of projectives of $\vs{\cc}$, with
quasi-inverse (``from shriek'') $\nu_{!\to}$.  Let $!\to *$ denote the functor
$\nu_{\to *}\circ\nu_{!\to}$, and similarly for $*\to !$ (exchanging
$*$ and $!$ in the subscripts).  These functors are clearly additive,
and therefore give rise to
functors from $\ck(\cp)$ to $\ck(\ci)$ and back (that are
quasi-inverses).  They therefore give rise to $\delta$-functors on
$\cd^{\rm b}(\vs{\cc})$.
(Again, the Axiom of Choice is used to construct quasi-inverses of the
natural maps from $\ck^{\rm b}(\cp)$ and $\ck^{\rm b}(\ci)$ to
$\cd^{\rm b}(\vs{\cc})$; this conceptually simplifying use of the
Axiom of Choice can be avoided as discussed in
Section~\ref{sb:axiom of choice}.)
We alternatively denote $(*\to !)F$ by $F^{*\to !}$ and similarly
for $!\to *$.

We note that, using Section~\ref{sbsb:representative}, any two
$!\to *$ constructed (from two different ``from shriek''
quasi-inverses $\nu_{!\to}$)
are isomorphic.  Similarly for $*\to !$.

Let $F\in\simexp{\cc}{\cv}$, and let $f\in\Hom(F,F)$.  We define
the trace of $f$ to be
$$
\trace(f) = \sum_{X\in\objects{\cc}} \trace\bigl(f(\id_X)\bigr).
$$
where 
$$
f(\id_X)\from F(X)\to F(X)
$$
is the restriction of $f$ to the identity morphism
on $X$, which is a linear map
from $F(X)$ to itself and therefore has a trace.

We claim that this trace is invariant under conjugacy, i.e., that if
$\iota\from F\to G$ is an isomorphism in $\simexp{\cc}{\cv}$, then
$\trace(\iota f \iota^{-1})=\trace(f)$.
To see this first note that for any $\delta\in\morphisms{\cc}$ we have
$$
\sum_{\gamma\alpha=\delta} \iota^{-1}(\gamma)\iota(\alpha)
=\left\{\begin{array}{ll} \id_{F(X)} & \mbox{if $\delta=\id_X$ for some
$X$,} \\ 0 & \mbox{otherwise,} \end{array}\right.
$$
by definition of composition and since $\iota^{-1}\iota=\id_{s\iota}$.
So
$$
\sum_X \trace\bigl( (\iota f\iota^{-1})(\id_X) \bigr) = 
\sum_{X,\alpha\beta\gamma=\id_X} \trace\bigl( \iota(\alpha)f(\beta)
\iota^{-1}(\gamma) \bigr)
$$
which, since $\alpha\beta\gamma=
\id_{t\alpha}$ iff $\beta\gamma\alpha=\id_{t\beta}$,
$$
= 
\sum_{X,\beta\gamma\alpha=\id_X} \trace\bigl( f(\beta)
\iota^{-1}(\gamma) \iota(\alpha) \bigr) 
= \sum_{X} \trace\bigl( f(\id_X) \bigr).
$$
Thus $\trace(\iota f\iota^{-1})=\trace(f)$.

Next we extend the trace on $\Hom(I,I)$ for each $I\in\ci$.  To do
this we choose an $F\in\simexp{\cc}{\cv}$ and an
$\iota\in\morphisms{\ci}$ such that $\iota\from\nu_{\to *}(F)\to I$ is an
isomorphism.
Given $f\in\Hom(I,I)$, $\nu_{\to *}$ sets up a one-to-one correspondence
between $\Hom(F,F)$ and $\Hom(\nu_{\to *}F,\nu_{\to *}F)$, and if
$g$ is mapped to $\iota f \iota^{-1}$ we define
$$
\trace(f) = \trace(g).
$$
We claim this definition of trace is independent of the choice of $F$ and
$\iota$.  Indeed, let $F',\iota'$ be another such choice.  Then $F,F'$
are conjugate under the morphism that maps (under $\nu_{\to *}$) to
$\iota^{-1}\iota'$ and the $g'$ that maps to $\iota'f'(\iota'^{-1})$
is conjugate to $g$ under this map.  Therefore $\trace(g)=\trace(g')$.

Next consider $f\in \Hom(I^{*\to !},I)$ for $I\in\objects{\ci}$.
We have $\nu_{*\to}I=\mu I$; also
$\nu_{*\to}$ on morphisms involving $I$
is determined by the choice of an $\iota\from \nu_{\to *}\mu I\to I$.
We get
$$
\iota^{-1}f\in\Hom(\nu_{\to !}(F),\nu_{\to *}(F)).
$$
From the direct sum decompositions of $\nu_{\to !}(F),\nu_{\to *}(F)$
we get restrictions, for each $X\in\objects{\cc}$
$$
(\iota^{-1}f)|_{{\rm Id}(X)}\from (k_X)_!F(X)\to (k_X)_*F(X).
$$
Since plainly
$$
\Hom\bigl((k_X)_!V,(k_X)_*V\bigr)\isom \Hom(V,V)
$$
on which we have the trace defined, we can define
$$
\trace(\iota^{-1}f) = \sum_{X\in\objects{\cc}} \trace\left(
(\iota^{-1}f)|_{{\rm Id}(X)} \right).
$$
But $\iota^{-1}f$ on ${\rm Id}(X)$ is independent of $\iota$ (since 
the $\iota$ must agree with $\mu$ on diagonal parts of morphisms).
Thus this trace is independent of the choice of $\iota$ and we can
unabmiguously denote it $\trace(f)$.

If $f\in\Hom(P,P^{!\to *})$ for $P$ projective,
we can similarly define the trace of $f$.

\section{Ext Duality}

Let ${\cal C}$ be a finite semitopological category.  Let
${\cal D}={\cal D}^{\rm b}(\myfield(\cc))$
be, as usual, the derived category of bounded $\myfield(\cc)$ complexes.
For $G\in {\cal D}$ we define a functor, $G^\LR$,
$$
F \mapsto \bigl( {\rm Hom}_{\cal D}(G,F) \bigr)^*.
$$
In this section we prove the following theorem.
\begin{theorem}\label{th:lefttoright}
For every $G$, $G^\LR$ is representable by
$G^{!\to *}$.
\end{theorem}
We call $G^\LR$ a the {\em left to right Ext dual of $G$}, for the
following reason.
\begin{corollary} Let $G^{!\to *}\isom H[n]$ in ${\cal D}$, where
$G,H\in\vs{\cal C}$.  Then for each $F\in\vs{\cal C}$ we have
$$
{\rm Ext}(F,H)= {\rm Ext}^n(G,F).
$$
\end{corollary}
In the case of the corollary above we say that $H$ is the {\em $n$-dimensional
left to right Ext dual of $F$}.
As mentioned before, $\LR$ or $!\to *$ is also known as the
{\em Serre functor}.
\begin{corollary} If $F^{!\to *}=F[n]$ and $G^{!\to *}=G'[n']$
for sheaves $G,G'\in\vs{\cal C}$ and integers $n,n'$ then
$$
{\rm cc}(F,G)={\rm cc}(F,G').
$$
\end{corollary}
The corollary follows since
$$
 \dim\bigl( {\rm Ext}^i(F,G) \bigr)
= \dim\bigl( {\rm Hom}_{{\cal D}}(F,G[i]) \bigr)
$$
$$
= \dim\bigl( {\rm Hom}_{{\cal D}}(G[i],F[n]) \bigr)
= \dim\bigl( {\rm Hom}_{{\cal D}}(F[n],G'[i+n']) \bigr).
$$
We finish this subsection with the proof of Theorem~\ref{th:lefttoright}.

First, if $A\in{\cal D}([0])$ we easily see that
$$
{\rm Hom}_{{\cal K}}(\myfield,A) \isom H^0(A),
$$
and
$$
{\rm Hom}_{{\cal K}}(A,\myfield) \isom \bigl( H^0(A) \bigr)^*.
$$
Next for $A\in\objects{{\cal C}}$, consider $k_A\from [0]\to {\cal C}$ as 
before.  Let $B\in{\cal D}^{\rm b}(\myfield({\cal C}))$.  
Then, using the fact that
$(k_A)_!\myfield$ is projective,
$$
{\rm Hom}_{\cal D}((k_A)_!\myfield,B) \isom 
{\rm Hom}_{{\cal K}({\cal C})}((k_A)_!\myfield,B) \isom 
{\rm Hom}_{{\cal K}([0])}(\myfield,k_A^*B) \isom H^0(k_A^*B),
$$
and similarly
$$
{\rm Hom}_{\cal D}(B,(k_A)_*\myfield) \isom  \bigl( H^0(k_A^*B) \bigr)^*.
$$
This shows that for {\em fixed} $A$, the functor $((k_A)_!\myfield)^{\LR}$
is represented by $(k_A)_*\myfield$.

\begin{definition} Consider a functor
$F\from {\cal T}\to \vs{\cal A}$, where ${\cal T}$ is a triangulated category
and ${\cal A}$ is an additive category.  We say that $F$ is a
{\em weak $\delta$-functor} if $F$ is additive and
for every distinguished triangle,
$X\to Y\to Z\to X[1]$ in ${\cal T}$ and every $A\in{\cal A}$ we have
$$
\cdots\to {\rm Hom}(FX[i],A) \to {\rm Hom}(FY[i],A) \to
$$
$$
{\rm Hom}(FZ[i],A) \to {\rm Hom}(FX[i+1],A) \to \cdots
$$
is exact.
\end{definition}
For example, let ${\cal A}$ be a triangulated category and
$F'\from{\cal T}\to{\cal A}$ a $\delta$-functor.  Then $F'$ followed
by the (vector space) Yoneda embedding is clearly a weak $\delta$-functor.

\begin{definition} A subset of objects, $I$, of a triangulated category,
${\cal T}$, is {\em triangularly closed} 
if for each distinguished
triangle $T_1\to T_2\to T_3\to T_1[1]$ in ${\cal T}$,
if any two of $T_1,T_2,T_3$ lie in $I$, then so does the third.
The {\em triangular closure} of a set of objects, $I$, in ${\cal T}$
is the intersection
of all triangularly closed sets in ${\cal T}$; we say $I$ 
{\em triangularly generates ${\cal T}$} if its triangular closure 
consists of all objects in ${\cal T}$.
\end{definition}

\begin{theorem} Let $F,G$ be two weak $\delta$-functors from ${\cal T}$
to $\vs{\cal A}$, and let $u\from F\to G$ be a morphism of functors.
Then the set of objects of ${\cal T}$ on which $u$ is an isomorphism
is triangularly closed.
\end{theorem}
\proof This follows from the
five-lemma (compare \cite{residues}, Proposition I.7.1).
\proofbox

\begin{theorem} Let $T$ be a subset of objects of an abelian category,
${\cal A}$, such that every object of $A$ has a finite resolution whose objects
are finite sums of elements in $T$.  Then $T$ triangularly generates
${\cal D}^{\rm b}({\cal A})$.
\end{theorem}
\proof This follows from \cite{residues}, Lemma I.7.2 (a distinguished
triangle in ${\cal D}({\cal A})$ consisting of an arbitrary middle element
and ``truncations'' from above and below on either side; also called
``filtrations'' in \cite{gelfand}, III.7.5), and from the fact that
if $A,B\in\ca$ then there is a distinguished triangle
$$
A\to A\oplus B\to B\to A[1]
$$
\proofbox

Next we wish to define a functor, $u$, from $!\to *$ to $\LR$, and to
show (1) $u$ is an isomorphism on each $(k_A)_*$, and then conclude
(2) therefore $u$ is an isomorphism on each object.

Let $P_\bullet$
be an object in the category of chains of projective sheaves,
${\rm Kom}(\cp)$.
For a chain map, $u\from P_\bullet\to(P_\bullet)^{!\to *}$, we define
$$
\trace(u) = \sum_i (-1)^i \trace(u_i),
$$
where $u_i$ is the map from $P_i$ to $(P_i)^{!\to *}$.

\begin{theorem} The trace of $u$ as above is independent of homotopy
class.  In particular, it gives rise to trace for each map from an
$F\in \cd^{\rm b}(\vs{\cc})$ to $F^{!\to *}$.
\end{theorem}
\proof 
It suffices to show this on each $X$-diagonal part, $X\in\objects{\cc}$.
In this case it suffices to show that for a map of finite dimensional
vector spaces $d\from V_1\to V_0$ and $K\from V_0\to V_1$ that
$\trace(Kd)=\trace(dK)$.  But this is a standard fact about traces.
\proofbox

Given two morphisms in $\cd^{\rm b}(\vs{\cc})$,
$f\from P_\bullet\to F_\bullet$ and $g\from F_\bullet\to P^{!\to *}$,
we have $\trace(gf)\in\myfield$; hence we have a map
$$
u\from \Hom(F_\bullet,P^{!\to *}) \to \bigl( \Hom(P_\bullet,F_\bullet) \bigr)^*
$$
defined for each $F_\bullet$.  
\begin{theorem} The above map $u$ is natural in both $P_\bullet$ and $F_\bullet$.
\end{theorem}
\proof
To see
naturality in $F_\bullet$, consider a map $g_\bullet\from F_\bullet\to G_\bullet$;
for each $w\in\Hom(G_\bullet,P_\bullet^{!\to *})$ and
$v\in\Hom(P_\bullet,F_\bullet)$
we have
$$
gv\in\Hom(P_\bullet,G_\bullet),\qquad
wg\in\Hom(F_\bullet,P_\bullet^{!\to *}),
$$
and naturality amounts to
$$
\trace\bigl( w(gv) \bigr) = \trace\bigl( (wg)v \bigr),
$$
which is clear.
Similarly, naturality in $P_\bullet$ reduces
to the fact that
for each $g\from P_\bullet\to Q_\bullet$,
$w\in\Hom(F_\bullet,P_\bullet^{!\to *})$, and $v\in\Hom(Q_\bullet,F_\bullet)$
we have
$$
\trace\bigl( w(vg) \bigr) = \trace\bigl( (g^{!\to *}w)v \bigr).
$$
This follows by applying $\mu$, whereupon $g$ and $g^{!\to *}$
become identified and the trace identity is standard.
\proofbox

Next we show that $u$ gives an isomorphism on objects for the
form $(k_A)_!\myfield$.  Indeed, we have already seen that
for $B\in{\cal D}^{\rm b}(\myfield({\cal C}))$,
$$
{\rm Hom}_{\cal D}((k_A)_!\myfield,B) \isom H^0(k_A^*B),
$$
and
$$
{\rm Hom}_{\cal D}(B,(k_A)_*\myfield) \isom  \bigl( H^0(k_A^*B) \bigr)^*.
$$
If $f\in\Hom_{\cal D}((k_A)_!\myfield,B)$ and $f\ne 0$, then
restricted to $A$, $f$ maps $1$ to a nonzero element of $H^0(k_A^*B)$.
Thus there exists a linear map $\ell\from H^0(k_A^*B)\to\myfield$ such
that $\ell(f(1))\ne 0$.  It follows that $\ell$ gives rise to
$g\in {\rm Hom}_{\cal D}(B,(k_A)_*\myfield)$ such that
$$
\trace(gf)=\ell\bigl(f|_A(1)\bigr)\ne 0.
$$
So $u$ for $(k_A)_!\myfield$
gives a morphism of vector spaces of the same dimension
that has a zero nullspace; thus $u$ is an isomorphism on each
$(k_A)_!\myfield$.

Since the elements of the form $(k_A)_!\myfield$
triangularly generate $\cd^{\rm b}(\myfield(\cc))$, 
we conclude the following theorem.
\begin{theorem} The above functor $u$ gives an isomophism of functors
$!\to *$ (followed by vector space Yoneda) with $\LR$.
\end{theorem}

\section{Local Criterion}

We are interested to know for which finite semitopological categories
we have $\myfield^{!\to *}\isom\myfield[n]$ for an integer $n$,
or $F^{!\to *}\isom F[n]$ for some $F$, and similar such conditions.
Here we give a necessary (but not sufficient) condition involving a
sort of local Euler characteristic.

Let $\cc$ be a finite semitopological category with adjacency matrix
$M$, i.e., the matrix indexed on pairs of objects of $\cc$ such
that $M(X,Y)$
is the size of $\Hom(X,Y)$.  
For an $F^\bullet\in\cd^{\rm b}(\vs{\cc})$ we define $v_F$ to be the
vector indexed on $\objects{\cc}$ whose $X$ component is
$$
v_F(X) = \sum_i (-1)^i \dim(F^i|_X).
$$
It is well defined on $\cd^{\rm b}(\myfield(\cc))$ and we shall call
it the local Euler characteristic.
Of course, $v_{F[n]}=(-1)^n v_F$ for all $n$.
The following theorem gives a necessary condition for checking duality:
\begin{theorem} If $G=F^{!\to *}$, then
$$
v_G = M^{\rm T}M^{-1} v_F .
$$
\end{theorem}
In particular, if $F^{!\to *}=F[n]$ for some $n$, then $v_F$ is an
eigenvector of $M^{\rm T}M^{-1}$ with eigenvalue $\pm 1$.  Similarly,
if $(F^{!\to *})^{!\to *}=F[m]$, then $v_F$ is a linear combination of
eigenvectors with eigenvalues $\pm 1$ if $m$ is even, and $\pm i$ if
$m$ is odd.
\proof First let $F\isom P_\bullet$ with $P_i\isom\bigoplus_X k_{X!}V_{X,i}$.
Let $w_F$ be defined by
$$
w_F(X) = \sum_i (-1)^i \dim V_{X,i}.
$$
Since
$$
\dim (k_{Y!}V)(X) = (\dim V) \ \bigl|\Hom(X,Y)\bigr|,
$$ 
we have
$$
v_F(X) = \sum_i (-1)^i \sum_Y (\dim V_{Y,i}) \ \bigl|\Hom(X,Y)\bigr|
= \sum_Y w_F(Y)  \ \bigl|\Hom(X,Y)\bigr|.
$$
Hence $v_F= Mw_F$.  
Similarly
$$
v_G(X) = \sum_i (-1)^i \sum_Y (\dim V_{Y,i}) \ \bigl|\Hom(Y,X)\bigr|,
$$
and we conclude $v_G=M^T w_F$.  Thus $v_G=M^T w_F=M^T(M^{-1}v_F)$.
\proofbox

\section{Strongly $n$-dimensional morphisms}

\begin{definition}\label{de:strongly}
Let $\phi\from\cc\to\cs$ be a morphism of semitopological
categories.  We say that $\phi$ is {\em strongly $n$-dimensional} if
$(!\to *)\phi^*$ is isomorphic to $[n]\phi^*(!\to *)$ as functors from
$\cd^{\rm b}(\vs{\cc})$ to $\cd^{\rm b}(\vs{\cs})$
\end{definition}

The point of this section and the next
is to prove that the above notion is stable
under base change.  We will also give alternative descriptions of
strong $n$-dimensionality.
First we note some easy alternative descriptions.

\begin{theorem}\label{th:strongly}
The following are equivalent:
\begin{enumerate}
\item $\phi$ is strongly $n$-dimensional, i.e.,
$(!\to *)\phi^*\isom [n]\phi^*(!\to *)$,
\item $\phi^*(*\to !)\isom [n](*\to !)\phi^*$,
\item $\lder\phi_!\isom [n]\rder\phi_*$.
\end{enumerate}
\end{theorem}
\proof Condition
(1) implies (2) by applying $*\to !$ and to left
and right sides; similarly (2) implies (1).  For condition (3) we
have that (1) implies
$$
\Hom(F_\bullet,[n]\phi^*(!\to *)P_\bullet) \isom \Hom(F_\bullet,(!\to *)\phi^* P_\bullet)
$$
so
$$
\Hom([-n]\lder\phi_! F_\bullet,(!\to *)P_\bullet) \isom \Hom(F_\bullet,(!\to *)\phi^* P_\bullet),
$$
so
$$
\Hom(P_\bullet,[-n]\lder\phi_! F_\bullet) \isom \Hom(\phi^* P_\bullet,F_\bullet)
\isom H(P_\bullet,\rder\phi_*F_\bullet),
$$
and vice versa.  Yoneda's lemma gives the desired
isomophism of functors.
\proofbox

\begin{definition} We say that $\cc$ is strongly $n$-dimensional if the
map from $\cc$ to $\cat{0}$ is strongly $n$-dimensional, or, equivalently,
if $\myfield^{!\to *}\isom \myfield[n]$.
\end{definition}

We mention that a number of models of $n$-dimensional manifolds are
strongly $n$-dimensional, not all of them are.  For example, let
$\cc$ be any strongly $n$-dimensional category, and let $\cc'$ be
the category obtained by adding one object, $X$, and one morphism to a
minimal element, $Y$, (and from each morphism from $Y$ adding one
corresponding
morphism from $X$).  If $\cc$ were obtained from a good cover of
a manifold, $\cc'$ could be obtained by adding one small neighborhood
of a point in a minimal open set.  It is easy to see that $\cc'$
will not be strongly $n$-dimensional, and indeed, that $(!\to *)\myfield$
restricted to $X$ is zero.

On the other hand, we do know a few examples of strongly $n$-dimensional
categories and morphisms.  The categories of $(\cat{1})^n$ with the
two extreme objects removed (modelling a good cover of $S^{n-2}$)
is strongly $(n-2)$-dimensional.  The category of $n+1$ pairs of
objects, modelling a successive upper and lower hemisphere covering
of $S^n$, and its quotient by the cyclic group of order two
are strongly $n$-dimensional.  Any Galois
morphism or covering morphism (i.e., a morphism that is Galois or
a covering space on the underlying graphs, as in \cite{geometric_aspects})
is strongly $0$-dimensional (it is immediate that such a morphism is fiberwise
$0$-dimensional, see below).  And (see below) any base change
of a strongly $n$-dimensional morphism is also one.  Hence any fiber product
of strongly dimensional
morphisms is one, and therefore so is any finite projective
limit involving only strongly dimensional morphisms
(see [SGA4.I.2.3.(iii)]).

We mention that the ``Boolean cube'' $\cat{1}^n$ for $n\ge 1$
is not strongly
$m$-dimensional for any $m$, and in fact $(!\to *)^3=[n]\id$ there.

{\bf Question:}\ \ We mention that at present we know of no $\cc$ with a
vector bundle $\omega\not\isom\myfield$ such that $\myfield^{!\to *}=
\omega[n]$ for some $n$.  
Can this happen?
We also mention that in all examples of 
strong dimensionality that we know, the ``skeleton'' 
(i.e., the objects, $X$, where $k_{X*}V_X$ with $V_X\ne 0$ is a
summand of one of the injectives)
of the greedy $\myfield$
injective (or projective) resolution is self-dual, in that there is a
simple isomorphism from $\cc$ to $\cc^{\rm opp}$ that maps the skeleton
to itself. 
Is this necessary?

\subsection{Main Result}
We now state the main theorem in this section.
Consider a base change diagram (i.e., a Cartesian diagram, i.e., where
$\cx'=\cx\times_\cs\cs'$):
\begin{equation}\label{eq:base_change}
\begin{array}{ccc} \cx' & \xrightarrow{f'} 
& \cs' \\
g'\downarrow & &\downarrow g \\
\cx & \xrightarrow{f} & \cs \end{array}
\end{equation}
As usual, we say that {\em $f'$ is obtain from $f$ via base change with
respect to $g$}.

\begin{theorem}\label{th:arbitrary}
Strong $n$-dimensionality is closed under arbitrary
base change; i.e., if $f$ in the above is strongly $n$-dimensional,
then so is $f'$.
\end{theorem}

Our proof of this fact will be to give another characterization of
strong $n$-dimensionality that is easily seen to
be closed under arbitrary
base change.
We call this characterization ``fiberwise $n$-dimensionality,'' and
it characterizes strong $n$-dimensionality in local terms, in terms
of the fiber over objects and the fiber over morphisms.

\begin{definition}\label{de:local}
If $f\from\cx\to\cs$ is a functor on finite, semitopological
categories, then we say that $f$ is {\em fiberwise $n$-dimensional}
if for any base change diagram as in equation~(\ref{eq:base_change})
with $\cs'=\cat{1}$ we have that $f'$ is strongly $n$-dimensional.
\end{definition}
We claim that the notion of fiberwise $n$-dimensional is easily seen
to be stable under base change.  Indeed, in a diagram of base changes:
$$
\begin{CD}
\cx'' @>{h'}>> \cx' @>{g'}>> \cx \\
@V{f''}VV @V{f'}VV @VVfV \\
\cat{1} @>h>> \cs @>g>> \cs
\end{CD}
$$
where $f$ is fiberwise $n$-dimensional, $g\from\cs'\to\cs$ is arbitrary,
and $h\from\cat{1}\to\cs'$ is arbitrary.  Then $g\circ h$ maps
$\cat{1}$ to $\cs$, and so $g''$ is strongly $n$-dimensional.  Thus
$f'$ is fiberwise $n$-dimensional.

Let us give a corollary of the base change theorem.
\begin{corollary} For $i=1,\ldots,k$, let $f_i\from \cc^i\to\cs$ be
a strongly $n_i$-dimensional functor.  Then their fiber product is a
strongly $n_1\ldots n_k$-dimensional functor.
\end{corollary}
We prove this by induction on $k$, using the easy fact that if
$\phi_1,\phi_2$ are strongly $n_1$- and $n_2$- (respectively) dimensional
morphisms, then $\phi_2\circ\phi_1$ is a strongly $n_1n_2$-dimensional
morphism.

\section{Proof of the Base Change Theorem}

Here we prove the equivalence of strong $n$-dimensionality and
fiberwise $n$-dimensionality.  We also make some further remarks
on base change.

\subsection{Strong is Stable Under Fully Faithful Base Change}

\begin{theorem}\label{th:open_closed}
Let $f\from\cx\to\cs$ be a strongly $n$-dimensional
functor of semitopological categories.  Then if $g\from\cs'\to\cs$
is an open or closed inclusion, then the base change morphism
$f'\from\cx'\to\cs'$ is strongly $n$-dimensional.
\end{theorem}
\proof
Here is the idea.
Let
$g\from\cs'\to\cs$ be an open inclusion.
Consider (1) of Theorem~\ref{th:strongly}, i.e,, the definition of
$!\to *$; then $!\to *$ on 
$\cx'$ and $\cs'$ may be computed on $\cx$ and $\cs$ (extending sheaves
from $\cx',\cs'$ to sheaves on $\cx,\cs$ by extension by zero) and
restricting back to $\cx'$ and $\cs'$.  Strong $n$-dimensionality is
clearly preserved by this process of extending by zero and then restricting.
Thus $f'$ is strongly $n$-dimensional.

In more detail, we have $g_!$ and $g'_!$ are exact and fully faithful,
and the above shows
$$
(!\to *)_{\cx'} \isom (g')^* (!\to *)_{\cx}\; g'_!,\quad
(!\to *)_{\cs'} \isom g^* (!\to *)_{\cs}\; g_!
$$
(all functors on the appropriate derived category).  Furthermore
it is easy to see that
$g'_!(f')^*\isom f^* g_!$ since $g$ is an open inclusion.  Thus
$$
(!\to *)_{\cx'} (f')^* \isom (g')^* (!\to *)_{\cx}\; g'_!(f')^*
\isom (g')^* (!\to *)_{\cx}\; f^* g_!
$$
$$
\isom (g')^* [n] f^* (!\to *)_{\cs}\;g_!
\isom [n] (f')^* g^*(!\to *)_{\cs}\;g_! \isom [n] (f')^* (!\to *)_{\cs'}.
$$

Similarly condition (2) of 
Theorem~\ref{th:strongly} shows that strong dimensionality is invariant
under closed inclusion base change.
\proofbox

Along similar but more involved lines we shall
prove that strong $n$-dimensionality
is stable under fully faithful base change.
\begin{theorem} Let $f\from\cx\to\cs$ be a strongly $n$-dimensional
functor of semitopological categories.  Then if $g\from\cs'\to\cs$
is fully faithful, then the base change morphism
$f'\from\cx'\to\cs'$ is strongly $n$-dimensional.
\end{theorem}

\begin{lemma}\label{lm:extend}
Let $f\from\cx\to\cs$ be strongly $n$-dimensional.
For $S\in\objects{\cs}$, 
let $\cx_S$ denote the full subcategory of $\cx$ with objects $f^{-1}(S)$.
Then a projective resolution of $\myfield$ on $\cx_S$ extends naturally
to one of $f^*k_{S!}\myfield$ on $\cx$ in the following sense:
let $\cdots\to P_1\to P_0\to\myfield$
be a projective resolution of $\myfield$ on $\cx_S$
with $P_i$ a sum of $k_{X!}V_{X,i}$ over $X\in f^{-1}(S)$;
then the sums of $k_{X!}V_{X,i}$ on $\cx$ (i.e., where now
$k_X\from \cat{0}\to\cx$ rather than $\cat{0}\to\cx_S$ as before)
give a projective resolution,
$\tilde P_i$,
of $f^*k_{S!}\myfield$.
We have the analogous statement about an injective resolution of $\myfield$
on $\cx_S$ extending to one of $f^*k_{S*}\myfield$ on
$\cx$.
Finally, if $\tilde I^{n-i}$ is the sum of $k_{X*}V_{X,i}$ on $\cx$
(with $V_{X,i}$
as above), then we have an
injective resolution of $f^*k_{S*}\myfield\to \tilde I^0\to 
\tilde I^1\to\cdots$
\end{lemma}
\proof
Consider a projective resolution of $\myfield$ on $\cx_S$ as in the
statement of the theorem.
By Theorem~\ref{th:open_closed} we have $f$ restricted to $\cx_S$ is
strongly $n$-dimensional, and hence $I^j=\oplus_X k_{X*}V_{X,n-j}$
gives an injective resolution of $\myfield$ restricted to $\cx_S$.
We apply strong dimensionality, in the form $f^*(*\to !)\isom [n](*\to !)f^*$
to $k_{S*}\myfield$ to conclude
$$
f^*k_{S!}\myfield \isom [n](*\to !)f^* k_{S*}\myfield.
$$
To find a projective resolution of $f^*k_{S!}\myfield$, it suffices
to do so when $\cs$ is replaced by the smallest open set containing $S$
(and $\cx$ replaced by $f^{-1}$ of this open set).  So we may assume
that $S$ is a terminal object of $\cs$.  In this case $f^*k_{S*}\myfield$
is just $\myfield_{f^{-1}(S)}$ (i.e., $\myfield$ on the closed set
$f^{-1}(S)$ and zero elsewhere), in which case $\tilde I^j=\oplus_X
k_{X*}V_{X,n-j}$ viewed on $\cx$ is visibly an injective resolution
(each $\tilde I^j$ is zero outside the closed set $f^{-1}(S)$ and
restricts to $I^j$ on $f^{-1}(S)$).  We therefore have
$$
f^*k_{S!}\myfield\isom [n](*\to !) \tilde I^\bullet.
$$
But $(*\to !)\tilde I^j\isom k_{X!}V_{X,n-j}$, so $[n](*\to !)\tilde I^\bullet$
is isomorphic to the projective resolution $\tilde P_\bullet$.
Thus
$$
f^*k_{S!}\myfield\isom \tilde P_\bullet.
$$

The statement regarding extending injective resolutions works by reversing
arrows.
The last statement holds by using the extension of an injective resolution,
given that the $I^j$ as above give an injective resolution of $\myfield$
restricted to $\cs$.
\proofbox

\begin{lemma} Let $f$ be strongly $n$-dimensional and $g$ be fully
faithful in the cartesian diagram of equation~(\ref{eq:base_change}).
Then there are isomorphisms $f^*\rder g_*\to (\rder g'_*)f'^*$
and $(\lder g'_!)f'^*\to f^*\lder g_!$.
\end{lemma}
\proof
We claim that for any $Q\in \objects{\cs'}$ we have
$$
f^*\rder g_* k_{Q*}\myfield \isom (\rder g'_*)f'^* k_{Q*}\myfield.
$$
Indeed, $\rder g_* k_{Q*}\myfield \isom k_{g(Q)*}\myfield$, so
the left-hand-side of the above displayed equation is simply
$f^*k_{g(Q)*}\myfield$.  But an injective resolution of this sheaf,
by the previous lemma, is given by the extension of a resolution 
of $\myfield$ on $\cx_{g(Q)}$; since $g$ is fully faithful so is $g'$
and hence $g'$ restricts to an isomorphism $\cx'_Q\to \cx_{g(Q)}$,
which is to say it is isomorphic (by the previous lemma) to
the image under $\rder g'_*$ to the injective resolution extended
from that of $\myfield$ on $\cx'_Q$, which is the right-hand-side
of the above equation.

Next we claim that the adjunctive map $\id\to (\rder g'_*)g'^*$ applied
to any $H$ in the image of $\rder g'_*$ is an isomorphism.  
Indeed, let $\ch$ be subcategory of $\cd^{\rm b}(\myfield(\cx))$ triangularly
generated by $k_{S*}\myfield$ over $S$ in the image of $g'$.
Since $g'$ is fully faithful, we see that $g'\from\ch\to\cd^{\rm b}
(\myfield(\cx'))$
is fully faithful.  Furthermore $(\rder g'_*)k_{T*}\myfield$, for each
$T\in\objects{\cx'}$, is isomorphic to $k_{g'(T)*}\myfield$; thus
the image of $\rder$ lies in $\ch$.
It follows by the end of Section~\ref{sb:adjoint} that 
$\id\to(\rder g'_*)g'^*$ is an isomorphism on the objects of $\ch$.

We know that $f^*\rder g_*k_{Q*}\myfield$ is in the image of $\rder g'_*$
for all $Q\in\objects{\cs'}$.  Thus the morphism obtained from 
the adjunctive morphism,
$$
f^*\rder g_*\to (\rder g'_*)g'^* f^*\rder g_*
$$
is an isomorphism on all $k_{Q*}\myfield$, and hence everywhere
(by closing triangularly).  Since $g$ is fully faithful and $fg'=gf'$,
we have
$$
g'^*f^*\rder g_* = f'^*g^*\rder g_* = f'^*,
$$
giving the isomorphism $f^*\rder g_*\isom
(\rder g'_*)f'^*$.

The other isomorphism comes from reversing the arrows. 
\proofbox

\begin{lemma} Let $g\from\cs'\to\cs$ be an arbitrary map.  Then
\begin{equation}\label{eq:star_shriek_swap}
\rder g_* (!\to *)_{\cs'} \isom (!\to *)_{\cs} \lder g_!
\end{equation}
Furthermore, in any diagram as in equation~(\ref{eq:base_change}), with
$f$ strongly $n$-dimensional, there is a canonical map
\begin{equation}\label{eq:automatic}
\mu\from
\rder g'_*[-n] (!\to *)_{\cx'}f'* \to (\rder g'_*) f'^*(!\to *)_{\cs'}
\end{equation}
once we give morphisms $(\lder g'_!)f'^*\to f^*\lder g_!$ and
$f^*\rder g_*\to (\rder g'_*)f'^*$; if the two given morphisms are 
isomorphisms, then so is $\mu$ above.
\end{lemma}

\proof For all $F\in\cd^{\rm b}(\myfield(\cs))$ and
$G\in\cd^{\rm b}(\myfield(\cs'))$ we have
$$
\Hom(F,(!\to *)(\lder g_!)G) \isom \Hom((\lder g_!)G,F) \isom
\Hom(G,g^*F)
$$
$$
\isom \Hom(g^*F,(!\to *)G)\isom \Hom(F,(\rder g_*)(!\to *)G)
$$
with the isomorphisms functorial in $F,G$.  This proves
equation~(\ref{eq:star_shriek_swap}).

For the second part have morphisms
$$
\rder g'_*[-n](!\to *)f'^*\to [-n](!\to *)(\lder g'_!)f'^* \to
[-n](!\to *)f^*\lder g_! 
$$
$$
\to f^*(!\to *)\lder g_!\to f^*\rder g_*(!\to *)\to (\rder g'_*)f'^*
(!\to *).
$$
\proofbox

Now we finish proving the theorem about stability of strong dimensionality
under fully faithful base change, by simply multiplying 
equation~\ref{eq:automatic} by $g'^*$ on the left and using that
$g'^*\rder g'_*\isom\id$.
\proofbox

\subsection{Strong is Stable Under Special Base Chage}

For integer $m\ge -1$, let $L_m$ be the category with two objects $0,1$ 
and with
$m+1$ morphisms from $0$ to $1$ (and no other nonidentity morphisms);
we call $L_m$ the {\em bouquet of $m$ loops}.
Of course, $\cat{1}\isom L_0$.
\begin{definition} By a {\em special functor}, we mean a functor
$u\from\cat{1}\to L_m$ for some $m\ge 1$ such that the objects $0,1$
in $\cat{1}$ are mapped to the same in $L_m$.
\end{definition}

Let $\cs$ be an arbitrary category such that $\Hom_\cs(A,B)$ is a finite
set for all $A,B\in\objects{\cs}$.
Any functor $\cat{1}\to\cs$ factors essentially uniquely as a special
morphism followed by a fully faithful morphism.
In this subsection we show that strong $n$-dimensionality
is stable under special base change, and therefore any base change
from $\cat{1}$.

\begin{definition} We say that a functor $f\from\cx\to\cs$ has the
{\em target lifting property} if for each $\phi\in\morphisms{\cs}$
and for each $T\in\objects{\cx}$ such that $f(T)$ is the target of
$\phi$, there is a $\xi\in\morphisms{\cx}$ whose target is $T$ and
such that $f(\xi)=\phi$.  The {\em source lifting property} is
similarly defined.
\end{definition}

\begin{theorem} Let $f\from\cx\to\cs$ be strongly $n$-dimensional.
Then $f$ has the source and target lifting property.
\end{theorem}
\proof Assume that $f\from\cx\to\cs$ is strongly $n$-dimensional.  Let
$S,T\in\objects{\cs}$, and let
$$
\Hom_\cs(S,T) = \{ \phi=\phi_1,\phi_2,\ldots,\phi_m \}.
$$
After an open inclusion and a
closed inclusion we may assume that $S$ is a minimal object of $\cs$ and
$T$ is a maximal one.  
Let
$\myfield|_{f^{-1}(S)}$ have projective resolution
of the form $P_\bullet$ with $P_i = \bigoplus_{f(X)=S} (k_X)_! V_{X,i}$.
Then $\myfield_{f^{-1}(S)}$ (the sheaf $\myfield$ on $f^{-1}(S)$ extended
by zero to the rest of $\cx$) has the same projective resolution as
$\myfield|_{f^{-1}(S)}$, provided the $(k_X)_! V_{X,i}$ are viewed on
$\myfield(\cx)$ (as opposed to just $\myfield(f^{-1}(S))$).
Hence
$$
(!\to *)f^*\myfield_S = (!\to *)\myfield_{f^{-1}(S)} = I^\bullet,
$$
where
$$
I^{-i} = \bigoplus_{f(X)=S} (k_X)_* V_{X,i}.
$$
On the other hand set
$$
F = f^* (!\to *) \myfield_S = f^*( (k_S)_*\myfield);
$$
for $Y\in\objects{\cx}$ and $\xi\from X\to Y$ with $f(X)=S$ we have
$$
F(Y) = \myfield^{\Hom(S,f(Y))},
$$
and $F\xi$ is zero on all components of $F(Y)$ except the one corresponding
to $f(\xi)$, on which it is the identity map to $F(X)=\myfield$.

First we claim that $f$ has the target lifting property.  
Indeed, consider the functor
$K\from\myfield(\cx)\to\myfield(f^{-1}(T))$
given by
\begin{equation}\label{eq:define_K}
(KG)(Y) = \bigcap_{i\ge 2,\,f(\xi)=\phi_i,\,{\rm targ}(\xi)=Y} \ker G\xi;
\end{equation}
$K$ is left exact, and so we have
$\rder K\from\cd^{\rm b}(\myfield(\cx))\to\cd^{\rm b}(\myfield(f^{-1}(T)))$.
Fix $Y\in\objects{\cx}$ with $f(Y)=T$.
On the one hand, it is easy to see that
if there is no $\xi$ with target $Y$ and
with $f(\xi)=\phi_1$, then
for any $X$ with $f(X)=S$
we have $(K(k_X)_*V_{X,i})(Y)$ vanishes. 
On the other hand,
we have $KF$ includes a copy of $\myfield$ (corresponding to $\phi_1$),
and hence $KF\not\isom 0$.  It follows that
$(\rder K)F\not\isom 0$ (since $H^0((\rder K)F)=KF$ since $F$ is a sheaf),
and hence $(\rder K)I^\bullet[-n]\not\isom 0$.
Thus $K$ cannot vanish at $Y$ when applied all the components $(k_X)_*V_{X,i}$
of $I^i$ for all $i$
(since $(\rder K)I^\bullet$ can be computed by applying $K$ to each $I^i$,
since each $I^i$ is injective).  Therefore $f$ has the target lifting property.

By symmetry, $f$ has the source lifting property.  
\proofbox

We mention that a strongly $n$-dimensional
$f$ need not be prefibered (see [SGA1.VI.6.1]).  Indeed, let $\cc,\cc'$
respectively be
copies of $L_1$, with objects $\{a,b\}$ and $\{a',b'\}$ respectively.
Consider
the union of $\cc$ and $\cc'$ with four additional arrows: one from $a$ or $b$ 
to $a'$ or
$b'$.  We call these four addition arrows ``a collection of zero arrows,''
since they do not affect the projective or injective resolutions of
$\myfield,\myfield_{\cc},\myfield_{\cc'}$.
Now take two copies of $\cc$ lying over a point, extend the base by
$\cat{1}$, to get a strongly $1$-dimensional $f'\from\cc''\to\cat{1}$;
$\cc''$ has two connected components.  Now take one component of
$f'^{-1}(0)$ and connect it to one of $f'^{-1}(1)$ via ``zero arrows'' as
above.  The resulting morphism, $f''$, is still strongly $1$-dimensional,
but is not prefibered, as there is no inverse image by $0\to 1$ of a point,
$P$, in
$f'^{-1}(1)$ in the component of the zero arrows (in particular, the 
category whose objects are the
morphisms over $0\to 1$ with target $P$ has no initial element).

Let us return to
a base change map as in equation~(\ref{eq:base_change})
with $\cs'=\cat{1}$, $\cs=L_m$,
$g(0)=0$, $g(1)=1$, $g(0\to 1)=\phi_1$, and
$\Hom_{\cs}(0,1)=\{\phi_1,\ldots,\phi_m\}$.

We take the functor $K$ of equation~(\ref{eq:define_K})
and extend it to all of $\myfield(\cx')$
by setting $\kbig\from\myfield(\cx)\to
\myfield(\cx')$ via
$$
(\kbig F)(X) = \left\{ \begin{array}{ll} F(g'(X)) & \mbox{if $f'(X)=0$,} \\
(KF)(X) & \mbox{if $f'(X)=1$.} \end{array} \right.
$$
We check that $\kbig$ extends to a functor (i.e., naturally acts on
$\myfield(\cx)$ morphisms).  Define
$Z_{\rm big}\from \myfield(\cx')\to\myfield(\cx)$ via $Z_{\rm big}F$
is $F$ on $g'(\cx')$ and $(Z_{\rm big}F)\phi_i=0$ for $i\ge 2$.
Define $\kbig'\from\myfield(\cx)\to
\myfield(\cx')$ via
$$
(\kbig' F)(X) = \left\{ \begin{array}{ll} K'(X) & \mbox{if $f'(X)=0$,} \\
F(g'(X)) & \mbox{if $f'(X)=1$,} \end{array} \right.
$$
where
$$
(K'F)(X) = F(g'(X))/\sum_{\xi{\rm \ s.t.\ }f(\xi)\ne\phi_1,s\xi=g'(X)}
{\rm image} F(\xi).
$$

We define $\ksmall,\zsmall,\ksmall'$ to be $\kbig,\zbig,\kbig'$ (respectively)
in the case where $f=\id$;  in other words,
$\ksmall,\ksmall'$ are maps $\myfield(\cs')\to\myfield(\cs)$
$$
(\ksmall F)(X) = \left\{ \begin{array}{ll} F(0) & \mbox{if $X=0$,} \\
\bigcap_{i\ge 2} \ker(F\phi_i) & \mbox{if $X=1$,} \end{array} \right.
$$
$$
(\ksmall' F)(X) = \left\{ \begin{array}{ll} F(0)/\sum_{i\ge 2}{\rm image}
(\phi_i) & \mbox{if $X=0$,} \\
F(1) & \mbox{if $X=1$,} \end{array} \right.
$$
and $\zsmall\from\myfield(\cs)\to\myfield(\cs')$ is given by
$(\zsmall F)(X)=F(X)$ for $X=0,1$, $(\zsmall F)(\phi)=F(0\to 1)$, and
$(\zsmall F)(\phi_i)=0$ for $i\ge 2$.

\begin{theorem} We have $\ksmall',\zsmall,\ksmall$ and $\kbig',\zbig,\kbig$
are sequences of adjoints.  
In particular $\lder\ksmall'=(*\to !)\rder\ksmall(!\to *)$ and similarly for
$\kbig'$ and $\kbig$.
Also $\kbig f^* = f'^*\ksmall$ and
$\kbig' f^* = f'^*\ksmall'$.
Finally, $\kbig g'_*\isom\id$, $(\rder \kbig)(\rder g'_*)\isom\id$,
and similarly for $\kbig'$ and $g'_!$ (and $\lder$ replacing $\rder$),
and similarly for $\ksmall,g$ replacing $\kbig,g'$ (respectively).
\end{theorem}
\proof
The first sentence (about adjointness) is a simple calculation we mostly
leave to the reader;  as an example, if $\mu\in \Hom(\zbig F,G)$ and
$v\in F(X)$ with $X\in f^{-1}(1)$, then since $(\zbig F)\xi=0$ for all
$\xi$ with $f(\xi)\ne\phi_1$,
we have $\mu(X)v\in\ker(G\xi)$ for all $\xi$ with $f(\xi)\ne\phi$, 
and as 
such $\mu$ gives rise to an element of $\Hom(F,\kbig G)$; similarly there
is an inverse map from $\Hom(F,\kbig G)$ to $\Hom(\zbig F,G)$, and similarly
a bijection $\Hom(\kbig' F,G)\to\Hom(F,\zbig G)$ (of course, the case
$\ksmall',\zsmall,\ksmall$ is a special case of $\kbig',\zbig,\kbig$).

The second sentence follows from equation~(\ref{eq:lr_adjoint}) using
the adjointness of the first sentence.

The third sentence follows almost immediately from
the fact that $f$ has the
target lifting property (for the first equation) and source lifting
(for the second).

In the fourth sentence, the case with $\ksmall,g$ replacing $\kbig,g'$
is just a special case.  So we need only deal with the
$\kbig,g'$ case.  That $\kbig g'_*=\id$ follows from the target lifting
property of $f$.  Since $g'_*$ takes injectives to injectives (since it
has an exact left adjoint), we have $\rder(\kbig)\rder g'_*=\rder(
\kbig g'_*)=\id$.  Similarly for $\kbig' g'_!=\id$ and
$\lder(\kbig')\lder g'_!=\id$.
\proofbox

\begin{theorem} We have $\rder(\kbig f^*)=(\rder\kbig)f^*$.
\end{theorem}
\proof It suffices to see that if $Y=0$ or $Y=1$, then $\kbig$ is
exact on the injective resolution of $f^*k_{Y*}\myfield$ (in other
words, $f^*k_{Y*}\myfield$ is $\kbig$-acyclic).  But the injective
resolution of $f^*k_{Y*}\myfield$ consists of components of the
form $k_{X*}\myfield$ for $X$'s with $f(X)=Y$, using
Lemma~\ref{lm:extend}.  Furthermore, $\kbig k_{X*}\myfield\isom k_{\tilde X*}
\myfield$ for the unique $\tilde X$ with $g'(\tilde X)=X$, and
$\kbig$ becomes an equivalence of categories when restricted to the
category triangularly generated by the $k_{X*}\myfield$ for
$X$ with $f(X)=Y$; this follows from the fact that if $f(x_1)=f(x_2)$
then
$$
\Hom_{\myfield(\cx')}(k_{\tilde X_1 *}\myfield,k_{\tilde X_2 *}\myfield)
\isom \bigl( \Hom(\myfield,\myfield) \bigr)^{\Hom_{\cx'}(\tilde X_1,
\tilde X_2)}
$$
$$
\isom \bigl( \Hom(\myfield,\myfield) \bigr)^{\Hom_{\cx}(X_1,X_2)}\isom
\Hom_{\myfield(\cx)}(k_{X_1 *}\myfield,k_{X_2 *}\myfield)
$$
for $X_i=g'(\tilde X_i)$.
Hence the desired exactness (or acyclicity) of $\kbig$.
\proofbox
We remark that while $\kbig$ is an equivalence of categories when restricted
to the subcategory triangularly generated by $k_{X*}\myfield$ over $X$ with
either $f(X)=0$ or $f(X)=1$, $\kbig$ is not generally right exact.  Here
is a simple example.
Consider on $\myfield(L_1)$ the surjection $F\to G$ where
$F=k_{0*}\myfield$, $G=k_{1*}\myfield$, and where $F\phi_1(a,b)=a$,
$F\phi_2(a,b)=b$, and $F(1)\to G(1)$ is given by $(a,b)\mapsto b$.
Then $F\to G$ is indeed a surjection, but $\kbig F\to \kbig G$ is the
zero map, not a surjection.

We now finish the special base change theorem.  We have
$$
\rder(\kbig)f^*\isom
\rder(\kbig f^*)\isom \rder(f'^*\ksmall)\isom f'^*(\rder\ksmall),
$$
the last equality since $\ksmall$ clearly maps injectives to injectives
($\ksmall k_{X*}\myfield\isom k_{X*}\myfield$ with $X=0$ or $X=1$, the
first $k_{X*}$ interpreted in $\myfield(\cs)$, the second in 
$\myfield(\cs')$). 
It follows that
$$
f'^* = \rder(\kbig)f^*\rder g_*.
$$
In the same way we conclude $f'^*=\lder(\kbig')f^*\lder g_!$.
Hence
$$
(!\to *)f'^* \isom (!\to *)\rder(\kbig)f^*\rder g_* \isom
\lder(\kbig')(!\to *)f^*\rder g_* 
$$
$$
\isom \lder(\kbig')[n]f^*(!\to *)\rder g_*
\isom \lder(\kbig')[n]f^*(\lder g_!) (!\to *)\isom [n]f'^*(!\to *).
$$

\subsection{Fiberwise implies strong}

Let $f\from\cx\to\cs$ be a fiberwise $n$-dimensional functor between
semitopological categories, and let
$$
\cdots\to\bigoplus_{X\in f^{-1}(S)} k_{X!}V_{X,1} \to
\bigoplus_{X\in f^{-1}(S)} k_{X!}V_{X,0} \to f^*k_{S!}\myfield
$$
be a fixed projective resolution of of $f^*k_{S!}\myfield$ for each
$S\in\objects{\cs}$ (which exists by Lemma~\ref{lm:extend}).
We may (and shall) assume that $(!\to *)f^*k_{S!}$ is the above projective
resolution with $k_{X!}$ replaced by $k_{X*}$; we also assume that
$(!\to *)_{\cs}k_{S!}V=k_{S*}V$ for all $S\in\objects{\cs}$.

We wish to exhibit
an isomorphism $\mu\from \nu_1\to \nu_2$, where
$\nu_1= (!\to *)f^*$ and $\nu_2=[n]f^*(!\to *)$.
Here is the overall strategy.  
Say that a $Q_\bullet\in\dercat{\cs}$ is {\em simple} if
\begin{equation}\label{eq:Q_i}
Q_i = \bigoplus_{X\in\objects{\cs}}(k_X)_! W_{X,i}
\end{equation}
for some vector spaces $V_{X,i}$.
First we will define $\mu$ on simple objects.  Second, we 
show that $\mu$ defines a natural transformation on the
full subcategory of whose objects are the simple ones.  Third, we extend
$\mu$ by general principles, using the fact that any element of the
derived category is isomorphic to a simple one, i.e., simple objects
are representative.

So fix a simple $Q$ as above and as in equation~(\ref{eq:Q_i}).
Consider the diagram below:
$$
\begin{CD}
Q_\bullet=\left(\oplus k_{S!} W_{S,i}\right)_{i\in\integers} @>{*\to !}>>
\left(\oplus k_{S*} W_{S,i}\right)_{i\in\integers} @>{[n]f^*}>>
\left(\oplus f^*k_{S*} W_{S,i+n}\right)_{i\in\integers} \\
@VV{f^*}V @. @A{\mu}AA \\
\left(\oplus f^*k_{S!} W_{S,i}\right)_{i\in\integers} 
@>{\isom}>>
\left(\oplus k_{X!} V_{X,j}\otimes W_{S,i}\right)_{i,j\in\integers}
@>{!\to *}>>
\left(\oplus k_{X*} V_{X,j}\otimes W_{S,i}\right)_{i,j\in\integers}
\end{CD}
$$
The arrow labelled $\isom$ arises since for any $S\in\objects{\cs}$
we have that $f^*k_{S!}\myfield$ has projective resolution
$$
\left( \bigoplus_{X\in f^{-1}(S)} k_{X!}V_{X,i} \right)_{i\in\integers}.
$$ 
But note that the composition of $\isom$ with $!\to *$ in the diagram
above
is just $!\to *$.
The two double complexes in the diagram
can be considered a single complex (and therefore
elements of the derived category) by the usual diagonal collapse.
We define $\mu$ via Lemma~\ref{lm:extend} and the isomorphism between
$$
\cdots\to
\bigoplus_{X\in f^{-1}(S)} k_{X*}V_{X,1} \to
\bigoplus_{X\in f^{-1}(S)} k_{X*}V_{X,0} \to 0 \quad\mbox{and}\quad
f^*k_{S*}\myfield.
$$

Let $\phi\from Q\to Q'$ be a map of simple objects.  We wish to
verify the commutativity of the diagram
\begin{equation}\label{eq:functorcommute}
\begin{CD}
(!\to *)f^* Q_1 @>{(!\to *)f^*\phi}>> (!\to *)f^* Q' \\
@V{\mu Q_1}VV @VV{\mu Q'}V \\
[n]f^*(!\to *) Q_1 @>{[n]f^*(!\to *)\phi}>> [n]f^* (!\to *) Q' \\
\end{CD}
\end{equation}
Let $S,T\in\objects{\cs}$ and $\psi\in\Hom(S,T)$, and consider
the special case $Q=k_{S!}\myfield$, $Q'=k_{T!}\myfield$;
since
$$
\Hom(Q,Q')\isom \bigl( \Hom(\myfield,\myfield) \bigr)^{\Hom(S,T)}
$$
(with a canonical isomorphism)
we set $\phi$ to $1^!_\psi$ which we define to be
the element of $\Hom(Q,Q')$ that is
(according to the right-hand-side of the equation displayed above)
the identity on $\psi$ and zero elsewhere.
Let's first look at
$$
(!\to *)f^*1^!_\psi .
$$
We have $f^* 1^!_\psi\from f^*k_{S!}\myfield\to f^*k_{T!}\myfield$,
and we get maps unique up to homotopy
\begin{equation}\label{eq:injective_map}
\begin{CD}
\cdots @>{d_2}>> \bigoplus_{X\in f^{-1}(S)} k_{X!}V_{X,1} @>{d_1}>> 
\bigoplus_{X\in f^{-1}(S)} k_{X!}V_{X,0} @>{d_0}>> f^*k_{S!}\myfield \\
@. @V{\gamma_1}VV @V{\gamma_0}VV @V{f^*1^!_\psi}VV \\
\cdots @>{d'_2}>> \bigoplus_{Y\in f^{-1}(T)} k_{Y!}V_{Y,1} @>{d'_1}>> 
\bigoplus_{Y\in f^{-1}(T)} k_{Y!}V_{Y,0} @>{d'_0}>> f^*k_{T!}\myfield \\
\end{CD}
\end{equation}
We claim that we may assume all the vertical arrows $\gamma_i$
``involve morphisms only over $\psi$'';
let us make this precise.
Recall that we construct $\gamma_0$, then $\gamma_1$, etc.\ using
projectivity; assume that we have chosen $\gamma_0$ via projectivity,
so that $d'_0\gamma_0=(f^*1^!_\psi)d_0$.
We have $f^* 1^!_\psi$ composed with $d_0$ determines an element of
$$
\Hom(k_{X!}V_{X,0},f^*k_{T!}\myfield)\isom
\bigl( \Hom(V_{X,0},\myfield) \bigr)^{\Hom(f(X),T)}
$$
for each $X\in f^{-1}(S)$
(of course, $f(X)=S$).
The application of $f^* 1^!_\psi$ here means this element is in the
image of $\psi$ from
$$
\bigl( \Hom(V_{X,0},\myfield) \bigr)^{\Hom(f(X),S)}.
$$
Since $\Hom(f(X),S)=\Hom(S,S)=\{\id_S\}$, we have that
$$
(f^*1^!_\psi)\circ d_0 \in \bigl( \Hom(V_{X,0},\myfield) \bigr)^{\Hom(f(X),T)}
$$
is zero on all $\Hom(f(X),T)=\Hom(S,T)$ components except possibly
$\psi\in\Hom(S,T)$.  This means that for any $Y\in f^{-1}(T)$, we have
can assume that $\gamma_0$ restricts to a map in
$$
\Hom(k_{X!}V_{X,0},k_{Y!}V_{Y,0})\isom
\bigl( \Hom(V_{X,0},V_{Y,0}) \bigr)^{\Hom(X,Y)}
$$
that is zero on all $\Hom(X,Y)$ components not in $f^{-1}(\psi)$
(by setting $\gamma_0$ to $0$ there and keeping $\gamma_0$ unchanged
on its $f^{-1}(\psi)$ components)
and obtain a commuting square $(f^*1^!_\psi)\circ d_0=d'_0\circ\gamma_0$.
Then $\gamma_0\circ d_1$ vanishes on $\Hom(X,Y)$ morphisms not in
$f^{-1}(\psi)$, and we may similarly assume $\gamma_1$ (chosen by projectivity
to satisfy $\gamma_0d_1=d_0\gamma_1$) has the same
property.  Similarly for all $\gamma_i$'s. 

The upshot is that equation~(\ref{eq:injective_map}) can be restricted
to $\cx'$ via
the base change $g\from\cs'=\cat{1}\to\cs$ with $g(0\to 1)=\psi$ without
loss of information.
So $(!\to *)f^*1^!_\psi$ is (without loss of generality) the map
\begin{equation}
\begin{CD}
\cdots @>{(!\to *)d_2}>> \bigoplus_{X\in f^{-1}(S)} k_{X*}V_{X,1} 
@>{(!\to *)d_1}>> 
\bigoplus_{X\in f^{-1}(S)} k_{X*}V_{X,0} @>>> 0 \\
@. @V{(!\to *)\gamma_1}VV @V{(!\to *)\gamma_0}VV @VVV \\
\cdots @>{(!\to *)d'_2}>> \bigoplus_{Y\in f^{-1}(T)} k_{Y*}V_{Y,1} 
@>{(!\to *)d'_1}>> 
\bigoplus_{Y\in f^{-1}(T)} k_{Y*}V_{Y,0} @>>> 0 \\
\end{CD}
\end{equation}
where in this diagram all arrows are supported on $f^{-1}(\psi)$ components.
Furthermore the other three arrows of equation~(\ref{eq:functorcommute}),
namely $\mu Q$, $\mu Q'$, and $[n]f^*(!\to *)1^!_\psi$, involve only
$f^{-1}(\psi)$ components, and the other two
objects of equation~(\ref{eq:functorcommute}), $[n]f^*k_{S*}\myfield$
and $[n]f^*k_{T*}\myfield$ (we may assume $(!\to *)k_{Y!}\myfield=k_{Y*}
\myfield$ for $Y\in\objects{\cs}$), are supported on $\cx'$.
So it suffices to verify the commutativity of
equation~(\ref{eq:functorcommute}) when viewed
on $\cx'$, (obtained by base change of
$f$ via $g$, i.e., restricting to ``$\psi$''),
but by fiberwise
$n$-dimensionality we have $f'$ is strongly $n$-dimensional, and
the commutativity is verified.


The case $Q=k_{S!}W$ and $Q'=k_{T!}W'$ follows similarly,
as does the case for arbitrary $\phi$, since any $\phi$ is a linear
combination of $1^!_\psi$'s.

We claim the general commutativity in equation~(\ref{eq:functorcommute}) 
now follows on all $Q,Q'$ each of whose members is a direct sum of
spaces $k_{S!}V$, since the morphisms in question operate componentwise
and decompose according to direct summands; let us write this out in
detail.
A morphism
$\phi\from Q\to Q'$ is a collection of morphisms $\phi_i\from Q_i\to
Q'_i$ (such that $d'_i\phi_i=\phi_{i-1}d_i$ for all $i$, where $d,d'$ are
the differentials of $Q,Q'$ respectively).  By assumption,
$$
Q_i=\bigoplus_{S\in \objects{\cs}} k_{S!}W_{S,i},\quad
Q_i=\bigoplus_{S\in \objects{\cs}} k_{S!}W'_{S,i}
$$
for each $i$, and so each $\phi_i$ is the direct sum of
$$
\phi_{i,S,S'}\in \Hom(k_{S!}W_{S,i},k_{S'!}W'_{S',i}).
$$
We wish to verify the commutativity of the diagram
\begin{equation}\label{eq:complicated_comm}
\begin{CD}
\bigoplus_{X,S=f(X)} k_{X*}(V_{X,j-n}\otimes W_{S,i})
@>{[-n](!\to *)f^*\phi}>> 
\bigoplus_{X,S=f(X)} k_{X*}(V_{X,j-n}\otimes W'_{S,i}) \\
@V{\mu Q}VV  @VV{\mu Q'}V  \\
\bigoplus_S f^*k_S W_{S,i} 
@>{f^*(!\to *)\phi}>>
\bigoplus_S f^*k_S W'_{S,i} \\
\end{CD}
\end{equation}
where the top row has double complexes viewed as complexes by, as usual,
grouping along the diagonals.  But
$$
(\mu Q')\circ \bigl( [-n](!\to *)f^*\phi \bigr)
$$
decomposes into components
$$
(\mu k_{S'} W'_{S',i})\circ \bigl( [-n](!\to *)f^*\phi_{i,S,S'} \bigr),
$$
and similarly for
$$
\bigl( f^*(!\to *)\phi \bigr) \circ (\mu Q)
$$
into components
$$
\bigl( f^*(!\to *)\phi_{i,S,S'} \bigr) \circ (\mu k_S W_{S,i}).
$$
But we have seen that the components agree,
hence the commutativity of equation~(\ref{eq:complicated_comm}).

Since every element of the derived category is isomorphic to a simple
$Q$,
we now appeal to general principles (see Paragraph~\ref{pa:nte}).
\proofbox

\section{The Base Change Morphism}

In this section we study the ``base change morphisms,'' natural
maps $g^*f_*\to f'_* g'^*$ and $f'_!g'^*\to g^*f_!$ studied, in the
context of the derived category, last section.
We believe such a study
may be useful in understanding various aspects of base change.

Consider a base change diagram (i.e., a Cartesian diagram):
$$
\begin{array}{ccc} \cx' & \xrightarrow{f'} 
& \cs' \\
g'\downarrow & &\downarrow g \\
\cx & \xrightarrow{f} & \cs \end{array}
$$

There is a morphism of functors, $u\from g^* f_*\to f'_*(g')^*$, known as
the {\em base change morphism}, described in [SGA4.XII.4] (page 6);
actually, two morphisms are described there, and later Deligne proves
([SGA4.XVII.2], with corrections to the proof given in 
[SGA4$\frac{1}{2}$], {\em Erratum pour SGA 4}) that they are the same morphism.
It is built from $f_*\to f_*g'_*(g')^*$ (from the adjunctive morphism)
which, since $f_*g'_*\isom g_*f'_*$ (canonically), gives a map
$f_*\to g_*f'_*(g')^*$ and therefore, by adjointness,
$g^*f_*\to f'_*(g')^*$.  

It is similarly possible to define a morphism $g^*\rder f_*\to
(\rder f'_*)(g')^*$.  We refer to this as the base change morphism for
the derived category.  If $(\rder f'_*)(g')^*=\rder( f'_*(g')^*)$,
then this morphism results directly from the base change 
morphism\footnote{It is not hard to see that 
$\rder(v_* u^*)$ is not generally isomorphic to $(\rder v_*)u^*$; indeed
take $v\from L_1\to\cat{0}$, 
and $u\from L_1\to\cat{1}$ (with $u$ an isomorphism on objects).  
Could it be the
case that we get an isomophism whenever $v=f'$ and $u=g'$ in a change
of base diagram, always or under some reasonable condition?
}.

We wish to know when the base change morphism is an isomorphim.
It turns out that it is under a number of interesting conditions, including
that $f$ is strongly dimensional, but it is certainly not always true.

First we shall study base change for sheaves of the form $k_{P*}\myfield$,
where $k_P$ is the one-point inclusion of $P\in\objects\cx$.

First note that the map $\id\to g'_*(g')^*$ of a sheaf $F$ is given by
the natural map
$$
F(X)\to \lim_{\substack{{\longleftarrow}\\{Z;g'(Z)\to X}}}F(g'(Z)).
$$
Since for $F=k_{P*}\myfield$ we have $F(Y)=\myfield^{\Hom(P,Y)}$, in
such a case the map $F\to g'_*(g')^*F$ is described as
\begin{equation}\label{eq:myfield_Hom}
\myfield^{\Hom(P,Y)}\to \myfield^M,
\end{equation}
where
$$
M=\lim_{\substack{{\longrightarrow}\\{Z;g'(Z)\to X}}} \Hom(P,g'(Z)),
$$
and where equation~(\ref{eq:myfield_Hom}) arises out of a set theoretic
map $L\to \Hom(P,Y)$.  Next we apply $f_*$, obtaining $f_*F\to
f_*g'_*(g')^*F\isom g_*f'_*(g')^*F$, and finally obtaining
$g^*f_*F\to f'_*(g')^*F$, each time only writing out the set theoretic maps
that give rise to these morphisms.  We easily see that the resulting set
theoretic map for $(g^*f_*F)(Q)\to (f'_*(g')^*F)(Q)$ is
$$
\lim_{\substack{{\longrightarrow}\\{W;f'(W)\to Q}}} \Hom(P,g'(W))
\to
\lim_{\substack{{\longrightarrow}\\{Z;g(Z)\to g(W)}}}
\lim_{\substack{{\longrightarrow}\\{W;f'(W)\to Q}}} \Hom(P,g'(W))
$$
$$
\to \lim_{\substack{{\longrightarrow}\\{W;gf'(W)\to Q}}} \Hom(P,g'(W))
\isom
\lim_{\substack{{\longrightarrow}\\{W;fg'(W)\to Q}}} \Hom(P,g'(W))
$$
$$
\to \Hom\bigl( f(P),g(Q) \bigr)
$$
So the surjectivity and injectivity (respectively) of the map
$(g^*f_*F)(Q)\to (f'_*(g')^*F)(Q)$ is equivalent to the injectivity and
surjectivity (respectively) of the natural map from
\begin{equation}\label{eq:base_limit}
w\from L = 
\lim_{\substack{{\longrightarrow}\\{W;f'(W)\to Q}}} \Hom(P,g'(W))
\to\Hom\bigl( f(P),g(Q) \bigr).
\end{equation}
This in turn amounts to factorizing morphisms $f(P)\to g(Q)$ by $f$ of a
morphism $P\to g'(W)$ followed by $g$ of a morphism $f'(W)\to Q$.
We introduce some terminology to make this precise.


\begin{definition}
If $P\in\objects{\cx}$ and $Q\in\objects{\cs'}$ 
then a $PQ$-morphism is a morphism $\phi\from f(P)\to
g(Q)$; we say that a {\em $PQ$-factorization of $\phi$} is a pair of morphisms
$\nu\in\morphisms{\cx}$ and $\mu\in\morphisms{\cs'}$ such that
$\nu$ has source $P$ and $\mu$ has target $Q$ and
$\phi=(g\mu)(f\nu)$.  We say that two $PQ$-factorizations of $\phi$,
$(\nu_1,\mu_1)$ and $(\nu_2,\mu_2)$ are primitively equivalent if
there exist $\nu_{12},\mu_{12}$ such that $\nu_2=\nu_{12}\nu_1$,
$\mu_1=\mu_2\mu_{12}$, and $f(\nu_{12})=g(\mu_{12})$.  We say that
$(\nu_1,\mu_1)$ and $(\nu_2,\mu_2)$ are equivalent if they are equivalent
under the transitive, reflexive closure of primitive equivalence.
\end{definition}

It is clear that in equation~(\ref{eq:base_limit}), a $PQ$-morphism, $\phi$, is
in the image of $w$ iff it has a $PQ$-factorization.  Furthermore, in this
case the preimage of $\phi$ will be unique iff all $PQ$-factorizations are
identified in the limit, $L$.
But the inductive limit over (a diagram of) sets is simply the disjoint
union of the sets modulo the equivalence relation that is the closure of
identifying two elements of two sets if one is mapped to the other in the
diagram.  Thus the preimage of $\phi$ will be unique iff any two
$PQ$-factorizations are equivalent.  We conclude the following theorem.

\begin{theorem} Consider the base change morphism
$u\from g^* f_*\to f'_*(g')^*$ described above.
Then $u$ applied to the sheaf $k_{P*}\myfield$ is injective
(respectively, surjective) at the object $Q\in\cs'$ iff
each $PQ$-morphism has a $PQ$-factorization (respectively, any two
$PQ$-morphisms are equivalent).
Hence $u$ is an isomorphism iff for each $P,Q$, every
$PQ$-morphism has a $PQ$-factorization and
any two $PQ$-factorizations of a morphism are equivalent.
\end{theorem}
(The last sentence makes use of the fact that the triangular closure of
the $k_{P*}\myfield$ is all of $\myfield(\cx)$.)

One can use this theorem to come up with conditions for which the base
change morphism is injective and/or surjective.  For example, if $f$ is
source liftable,
it is easy to see that
any $PQ$-morphism, $\phi$, has a $PQ$-factorization (where the second morphism
is ${\rm id}_Q$), and any $PQ$-factorization is equivalent to one where the
second morphism is ${\rm id}_Q$.
If $f$ is precofibered
(see [SGA1.VI.6.1]), then it is immediate 
that any two $PQ$-factorizations where
the second morphism is ${\rm id}_Q$ are equivalent; however, we imagine the
``unique up to equivalence'' condition of the above theorem to be much weaker
than that of being precofibered; precofibered implies that the 
$PQ$-factorizations with second morphism ${\rm id}_Q$ has a terminal element,
whereas ``unique up to equivalence'' speaks of the connectedness of a
(possibly larger) category.

To give an example of when the conditions of the theorem are not satisfied, 
take
$\cs=\cat{1}$ and $f,g$ to be the inclusions of $\cat{0}$ into $0$ and $1$
respectively (in this case $\cx'$ is the empty category).

\appendix
\section{Simple Remarks on Duality}
\label{ap:duality}

In this section we try to generalize the setting of Ext duality and
make some aspects of it more precise.  The
idea is that $\Hom$ of two objects of the category in question should
carry some extra structure (such as that of a finite dimensional
vector space) that has some notion of a dual.

In general there seem to be two approaches.  First, one can speak about
representability, adjoints, etc.\ with respect to the new structure of
$\Hom$, and hope this is compatible with the old (set theoretic)
structure.  This is what seems to be commonly done with Serre functors,
and this is what we have done in the paper (and works fine).  
Another method is to lay down
axioms regarding this new structure that ensure some level of
compatibility; the more structure, the more automatic the compatibility
becomes.  We shall do this in two subsections: first we give what seems like
minimal structure, but enough to talk about $\LR$ (or Serre) functors;
second we give enough structure that guarantees that representability
means essentially the same thing.

\subsection{Hom structures}

\begin{definition} Let $\cm,\cv$ be categories.  To give $\cm$
a {\em $\cv$-Hom} structure is to give a functor
$$
\Hom_\cm^\cv\from\cm^{\rm op}\times \cm\to\cv
$$
and a functor
$$
{\rm forget}\from\cv\to({\rm Ens})
$$
(where (Ens) is the category of sets or elements of a universe)
such that ${\rm forget}\circ\Hom_\cm^\cv\isom \Hom_\cm$.
\end{definition}
In the above case
one can assume that
${\rm forget}\circ\Hom_\cm^\cv = \Hom_\cm$ by redefining $\Hom_\cm$.

\begin{definition} A {\em reversible category}, $\cv$, is a category
that is a $\cv$-category 
with a contravariant functor, $\iota$, to itself that is
an essential involution, i.e., $\iota^2\isom\id$.
\end{definition}

Let $\cv$ be a reversible category, and $\cm$ a $\cv$-category.  For
each $B\in\objects{\cv}$ we define the {\em left-to-right} functor,
$B^{\LR}$ or $(\LR)B$ to be the functor
$$
A\mapsto ({\rm forget})\iota\Hom_\cm^\cv(B,A);
$$
as a function of $B$ this functor $B^{\LR}$ is clearly covariant functor.
If $B^{\LR}$ is representable for each $B$, then Yoneda's lemma shows that
$(\LR)$ gives a 
(covariant) functor from $\cm$ to itself; it is defined uniquely up
to (unique) isomorphism and it is ambiguous up to isomophism. 
By abuse of notation we use $B^{\LR}$ to denote the object representing
$B^{\LR}$.  We say that the left-to-right functor is 
{\em $\cv$-representable} if there
is an isomophism
$$
\Hom_\cm^\cv(A,B^{\LR})\isom \iota\Hom_\cm^\cv(B,A)
$$
that is natural in $A$.

The right-to-left functor is defined analogously, via the equation
$$
\Hom_\cm^\cv(A^{\RL},B)\isom \iota\Hom_\cm^\cv(B,A).
$$
If both right-to-left and left-to-right functors are $\cv$-representable, then
$\LR$ and $\RL$ are quasi-inverses of each other, since
$$
\Hom_\cm^\cv(A,B)\isom \iota\Hom_\cm^\cv(B,A^{\LR})
$$
$$
\isom\iota^2\Hom_\cm^\cv\bigl((A^{\LR})^{\RL},B\bigr)
\isom\Hom_\cm^\cv\bigl((A^{\LR})^{\RL},B\bigr),
$$
and we may apply Yoneda's lemma and the functor $({\rm forget})$.

Let $\cm_1,\cm_2$ be categories with
$\cv$-Hom structure with $\cv$ reversible.  Consider
a pair of functors, $(F,G)$, with $F\from \cm_1\to\cm_2$ and
$G\from\cm_2\to\cm_1$.  We say that $F$ is a left $\cv$-adjoint to $G$
(or $G$ a right $\cv$-adjoint to $F$) if there is an isomorphism
of bifunctors
$$
\Hom_{\cm_2}^\cv(FA,B) \isom \Hom_{\cm_1}^\cv(A,GB)
$$
in $A$ and $B$.
We easily see the following theorem (remarked in \cite{larsen}, for
vector spaces).

\begin{theorem}
\label{th:interesting}
Under the notation and assumptions of the previous paragraph,
we have that $G$ has a right adjoint
$$
(\LR)_{\cm_2} F (\RL)_{\cm_1},
$$
provided the above left-to-right and right-to-left functors $\cv$-exist.
\end{theorem}
\proof We have
$$
\Hom_{\cm_1}(GA,B)\isom \Hom_{\cm_1}\bigl( (\RL)_{\cm_1}B,GA \bigr) \isom
$$
$$
\Hom_{\cm_2}\bigl( F(\RL)_{\cm_1}B,A \bigr) \isom
\Hom_{\cm_2}\bigl( A, (\LR)_{\cm_2} F(\RL)_{\cm_1}B \bigr).
$$
\proofbox

We conclude a similar theorem about the left adjoint of $F$.  More generally
assume the left-to-right and right-to-left functors are $\cv$-representable
in both
$\cm_1$ and $\cm_2$.  Then there is a sequence of left/right adjoints
$$
\cdots,F_{-1},G_{-1},F=F_0,G_0,F_1,G_1,\cdots
$$
where
$$
F_i = (\LR)^i_{\cm_2} F (\RL)^i_{\cm_1}, \quad
G_i = (\RL)^{-i}_{\cm_1} G (\LR)^{-i}_{\cm_2},
$$
and where $(\LR),(\RL)$ taken to a negative exponent means taking
$(\RL),(\LR)$ respectively to the corresponding positive exponent.

\subsection{Representability}

We finish this section by trying to make the above discussion a bit
more satisfactory.  Namely, in the above we spoke of
$\cv$-representability, whereas in practice it should follow ``automatically''
from representablility; similarly adjoints should always be $\cv$-adjoints.
This is clear in the case used in this paper, where $\cv$ is the
category of finite dimensional vector spaces
and ``forget'' is the usual forgetful functor.
Yet, we'd like to convince
the reader that this automatic carrying over to the $\cv$-structure can be
done with some simple axioms that don't seem overly restrictive.

(Also note that 
$\cv$-representability and $\cv$-adjoints are all that are necessary to
the previous subsection, to discuss $\LR$ functors---
representability, adjoints, and the functor ``forget''
are not necessary, but give the standard application.)

We fix notation as in the previous subsection, 
with $\cm$ being a category with $\cv$-Hom
structure where $\cv$ is reversible.  We add the following axioms:
\begin{enumerate}
\item $\cv$ has a unit, $u$, i.e., an object such that
$$
\Hom_\cv^\cv(u,\;\cdot\;) \isom {\rm Id}_\cv
$$
and
$$
\Hom_\cv(u,\;\cdot\;) \isom {\rm forget};
$$
\item the operation $A\Box B=\Hom_\cv^\cv(\iota A,B)$ has an associated
natural isomorphism in $A,B,C$:
$$
(A\Box B)\Box C \isom A\Box (B\Box C)
$$
\item both categories $\cm,\cv$ have ``${\rm Hom}$ compositions''
$$
{\rm comp}_\cc^\cv \in \Hom_\cv^\cv\Bigl( \Hom_\cc^\cv(A,B),
\Hom_\cv^\cv\bigl(\Hom_\cc^\cv(B,C), \Hom_\cc^\cv(A,C) \bigr) \Bigr)  
$$
for $\cc=\cm,\cv$, natural in $A,B,C$, and ``identities''
$$
{\rm id}_{A,\cc}^\cv \in \Hom_\cv^\cv\bigl( u,\Hom_\cc^\cv(A,A)\bigr)
$$
that satisfy the usual category identities and that restrict to their
set-theoretic analogues upon applying the forgetful functor.
\item a $\cv$-morphism is a monomorphism iff it is after applying
``forget,'' and similarly for epimorphism and isomorphism.
\end{enumerate}

A few remarks on these conditions.  First, condition~(2) is equivalent
to
$$
\Hom_\cv^\cv(A\Box B, C) \isom 
\Hom_\cv^\cv\bigl( A,\Hom_\cv^\cv(B,C) \bigr),
$$
the usual ``exponential'' condition.  Note that some relations, such as
$A\Box B\isom B\Box A$ are automatic from the definition of $\iota$.
We don't know if condition~(2) is absolutely necessary, but it certainly
simplifies the discussion below; similarly for condition~(4).  
We can speak of passing
from a $\cv$-morphism
$$
\Hom_\cc^\cv(A,B)\Box \Hom_\cc^\cv(B,C) \rightarrow \Hom_\cc^\cv(A,C)
$$
to one of sets
$$
\Hom_\cc(A,B)\times \Hom_\cc(B,C) \rightarrow \Hom_\cc(A,C)
$$
by writing things as in condition~(3) and applying the functor
${\rm forget}$; similarly for any condition that can be written in
terms of $\Hom^\cv$'s and $u$'s.

We finish by showing that representability in $\cv$ or in (Ens) is
the same.  Clearly $\cv$-representability implies (Ens)-representability.

Now let $F$ be a contravariant $\cv$-functor, meaning
$F\from\cm^{\rm op}\to \cv$ such that the functoriality is given by a
map of vector spaces,
$$
\Hom_\cm^\cv(A,B)\Box F(B)\rightarrow F(A),
$$
for each $A,B$ (natural in $A,B$).  Assume that $({\rm forget})F$ is 
represented by $M$, i.e.,
$$
({\rm forget})F \isom \Hom_\cm(\;\cdot\;,M).
$$
Then we claim that $F$ is $\cv$-representable by $M$.  Indeed,
we have $\id_M$ corresponds to an element of $({\rm forget})F(M)$ that
corresponds to a $\cv$-morphism $u\to F(M)$.  We therefore get a map
$$
\Hom_\cm^\cv(A,M)\isom \Hom_\cm^\cv(A,M)\Box u \to \Hom_\cm^\cv(A,M)\Box F(M)
\to F(A).
$$
Upon applying ``forget'' we easily see that the above morphism
$\phi_A\from\Hom_\cm^\cv(A,M)\to F(A)$ 
maps to one direction of the isomophism between
$\Hom_\cm(A,M)$ and $({\rm forget})F(A)$.  
It follows that $\phi_A$ is an isomorphism for each $A$, and therefore
has a unique inverse, $\mu_A$.
This easily gives the desired $\cv$-representability.

\end{document}